\documentclass[12pt]{iopart}

\usepackage[style=gb7714-2015]{biblatex}
\usepackage{indentfirst}
\usepackage{amssymb}
\usepackage{graphicx}
\usepackage{float}
\usepackage{listings}
\expandafter\let\csname equation*\endcsname\relax
\expandafter\let\csname endequation*\endcsname\relax
\usepackage{amsmath}
\usepackage{subfig,graphicx}
\usepackage{caption}
\usepackage{epstopdf}
\usepackage{epsfig}
\usepackage{iopams}
\renewcommand{\thefigure}{\arabic{figure}}

\renewcommand{\thetable}{\Roman{table}}
\gdef\theequation{\arabic{section}.\arabic{equation}}
\usepackage[colorlinks,linkcolor=blue,anchorcolor=blue,citecolor=blue]{hyperref}

\begin{document}
\title{Modular Study of a Force-Magnetic Coupling System}

\author{Zifeng Li$^{a,b,*}$ , Yuanmei Li$^{a,b,}$\footnote[7]{First Author and Second Author contributr equally to this work.} , Yinlong Wang$^{a,b,c}$,Biao You$^{a,b}$ ,Jianguo Wan$^{a,b}$}
\address{$^a$National Demonstration Center For Experimental Physics Education, Nanjing University, 163 Xianlin Road, Qixia District, NANJING, Jiangsu 210023, China}
\address{$^b$School of Physics, Nanjing University, 22 Hankou Road, Gulou District, NANJING, Jiangsu 210023, China}
\address{$^c$email: wangylphy@nju.edu.cn; tel: +86 025 8539 6530}
\vspace{10pt}
\begin{indented}
\item[]August 2024
\end{indented}

\begin{abstract}

A magnetic-mechanical oscillating system consists of two identical leaf springs, a non-magnetic base, and some magnets. The leaf springs are fixed at the bottom to the non-magnetic base, while the magnet is attached to the top of the leaf springs. This paper investigates the overall motion characteristics of the magnetic-mechanical oscillating system.

Adopting the modular modeling concept, we simplify the system into three inter-coupled modules: the leaf springs, magnetic interactions, and the system's dissipation process. We conduct physical modeling and theoretical analysis on these modules and derived the system's dynamic equations. The research indicates that the system is a normal mode system with two degrees of freedom.

In addition, we alter parameters and conduct multiple innovative experiments, obtaining intuitive vibration images that characterize the vibration modes and the periodic energy transfer.

Furthermore, we employ the simulation software COMSOL Multiphysics simulation to substitute the theory for auxiliary validation, achieving a comprehensive research loop of theory-experiment-simulation. The experimental results show good consistency with the theoretical calculations and simulation results.

This research provides a good teaching case for magnetic-coupling complex systems. This modular analysis and rather practical experimental design could solve the previous difficulty that the solution to such problem is too complex, and is conducive to the implementation of education.

\end{abstract}
\noindent{\it Keywords}:\hspace{0.2cm} Oscillating, Magnetic Interactions, Normal Mode System, Modular Modeling
\newpage

\section{Introduction}

\subsection{Research background}

Multi-degree-of-freedom coupled harmonic oscillating system is a classic model that have attracted the interest of many researchers due to its simple composition and rich physical implications. This kind of system is still evolving, achieving new progresses with great research values:
Adya Wadhwa and Ajay Wadhwa proposed a simple and elegant pendulum experimental device for measuring the damping coefficient of vibrations. Using a simple pendulum as an example, they studied single-degree-of-freedom vibrations, which is a common example in physical education \cite{item1};
T.Corridoni and M.D'Anna showcased their skills in a regular two-degree-of-freedom coupled vibration system. They used the method of coordinate transformation to analyze its properties, achieving selective suppression between two vibration modes \cite{item2};
Qinghao Wen and Liu Yang provided a detailed and clear analysis of the behavior of the Huygens' pendulum, a classic two-degree-of-freedom vibration system \cite{item3};
In the study of multi-degree-of-freedom systems, James Pantaleone placed multiple metronomes on a freely moving base, showing that even randomly started metronomes eventually synchronize due to coupling \cite{item4}.

In the context of electromagnetism, magnetic interactions through magnetic fields between magnets are a common form of interaction. Magnetic forces are also popular teaching examples for educators:
Manuel I.González studied the forces between two parallel magnets in detail and proposed a simple and low-cost experimental device to measure these forces. This allows students at various learning stages to intuitively appreciate the wonders of electromagnetism \cite{item5};
The properties of mechanical systems centered around magnetic forces are also a highlight of experimental teaching innovations. These systems can intuitively demonstrate the special distance-dependent nature of magnetic interactions and the fascinating nonlinear phenomena \cite{item6,item7,item8,item9}.

Thus, combining these two appealing elements, many researchers and educators have conducted studies on complex systems involving mechanical and magnetic coupling or equivalent systems. \footnote[7]{Since the interaction mechanism and the functional expressions are similar, equating the magnetic interaction in this system to a spring is a commonly used and highly practical approach.}

Bernard J. Weigman and Helene F. Perry applied the Lagrange equations to directly solve the motion of a one-dimensional coupled system consisting of a spring and a mass block, and they obtained reliable experimental results using advanced measurement equipment and automated analysis methods\cite{item10};

Lance McCarthy studied a coupled magnetic-mechanical oscillator, fitting the transient response experimental results into differential equations derived from the system's properties, which preliminarily addressed the challenge of obtaining reliable quantitative theoretical results for such complex systems\cite{item11};

In addition to relatively low-degree-of-freedom coupled harmonic oscillatory systems, Ross L. Spencer and Richard D. Robertson investigated multi-degree-of-freedom magnetically coupled vibrating blade systems. Due to the weak coupling in this system, unusual vibration modes were observed. The authors employed a theoretical approach similar to that of the previous study, combined with extensive experiments to explore the detuning within this system\cite{item12};

Although the approaches mentioned above are highly ingenious and have, to some extent, addressed the challenges previously encountered in the analysis of such systems, the complexity of their theoretical methods has led to a gradual decline in interest in these physics-rich and educationally valuable studies on mechanically and magnetically coupled complex systems. As a result, there has been a noticeable decrease in related research in recent years, which we find quite regrettable.

In this article, we focus on studying the magnetic-mechanical oscillating system with the intent to clearly, intuitively, and rigorously showcase the characteristics and highlights of the entire system. Unlike previous approaches, we have adopted a modular method, analyzing and processing different parts of the system separately. By simplifying the complex problems and constructing an effective physical model, we have successfully tackled the challenge of analyzing force-magnetic coupling systems. More importantly, when analyzing the bending of a plate, many undergraduates may assume that the degree of curvature along the plate's length is linear, which they can use to analyze the plate's vibrational kinetic energy. However, this is not correct, even for a plate with minor deformations. The curvature along the plate's length is not a linear function, but a cubic function. We should be careful when handling the kinetic energy of the bending plate vibration process. Based on Young's modulus, we will give a detailed and quantitative analysis of the non-linear bending phenomenon of the plate.

\subsection{Research summary}

The magnetic-mechanical oscillating system is a two-degree-of-freedom vibration model coupled by magnetic forces. This mechanical system shares some properties with multi-degree-of-freedom systems and magnetic interactions: including approximately linear elastic restoring forces, nonlinear magnetic interactions, and amplitude-dependent vibration frequencies.

When dealing with this complex real-world problem, many factors need to be considered. In this article, we adopted a modular approach, modeling the magnetic-mechanical oscillator system as three independent modules: the vibration of the leaf springs, magnetic interactions, and the dissipation process, and handled them separately.

To implement the magnetic-mechanical oscillating system, we designed a structurally intuitive and low-cost experimental device. By simply and accurately changing parameters, we conducted multiple experiments and obtained aesthetically pleasing and intuitive vibration images.

Furthermore, we used CAE software COMSOL Multiphysics to assist in analyzing the properties of each module, ultimately completing the rigorous process of theory-experiment-simulation comparison in this research work. We proposed a dimensionless number $Z$ to determine the applicability of our theory.

Our experimental design is easy to reproduce and our theory has good applicability, which can innovate teaching methods by showing students the charm of coupled systems. This theoretical analysis and mechanical design can also inspire subsequent research work, contributing to theoretical physics or industrial research fields.

Further, our research is revolutionary and highly reality-relevant. In reality, most everyday phenomena are not isolated mechanical or electromagnetic systems; rather, they are often coupled together. Our approach offers a solution to the challenges posed by such coupled systems and serves as an excellent case study for university physics education.

\subsection{Brief understanding}

\textbf{Figure \ref{Fig.1}} depicts a schematic diagram of a magnetic-mechanical oscillator:

\begin{figure}[h]
\begin{minipage}[b]{0.68\linewidth}
    \centering
    \subfloat[A magnetic-mechanical oscillator]{\includegraphics[width=1\linewidth]{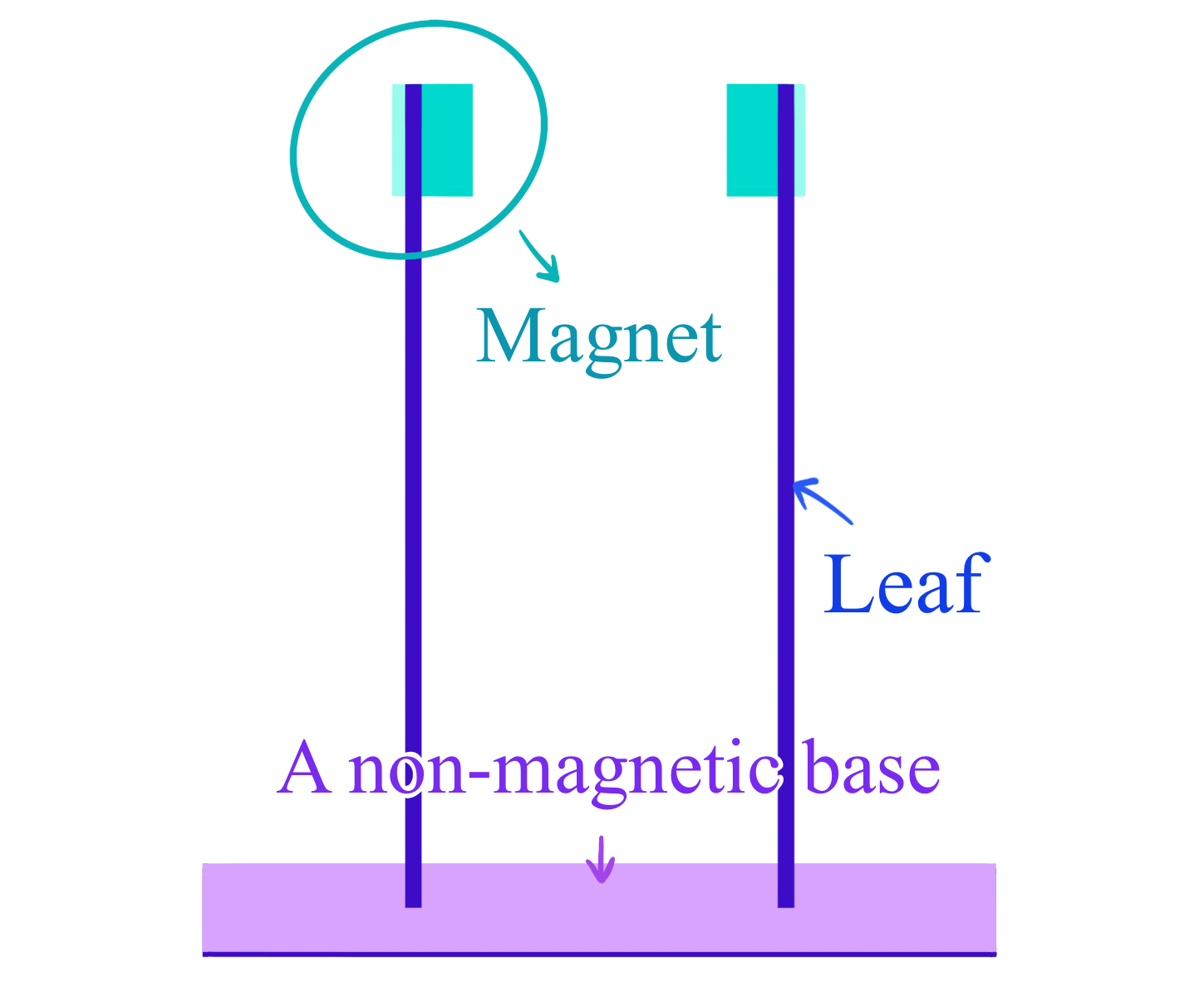}}
    \label{Fig.1(a)}

\end{minipage} 
\medskip
\begin{minipage}[b]{0.31\linewidth}
    \centering
    \subfloat[Oscillation of one
plate]{\label{Fig.1(b)}\includegraphics[width=1\linewidth]{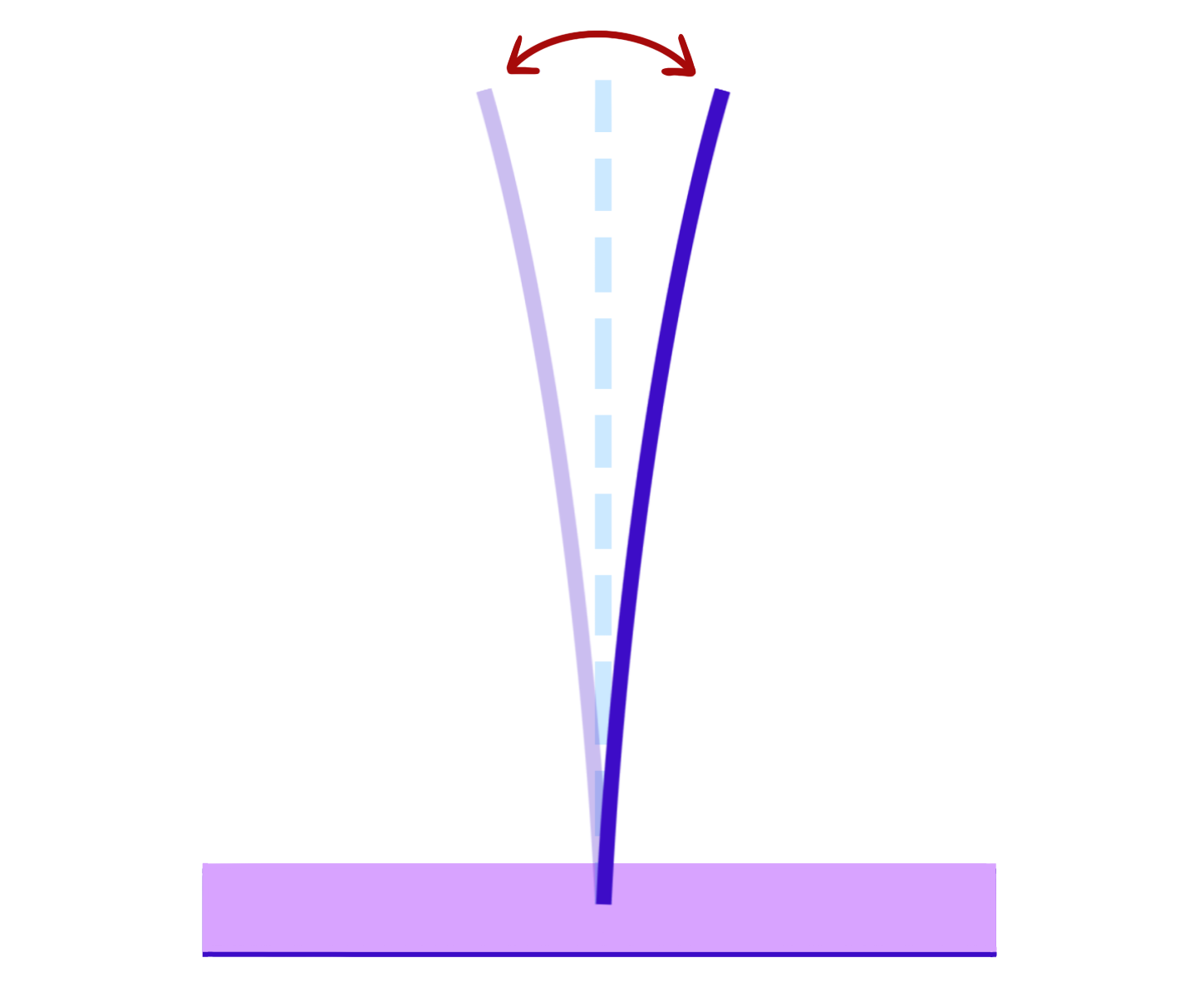}}

    \subfloat[Oscillation induction]{\label{Fig.1(c)}\includegraphics[width=1\linewidth]{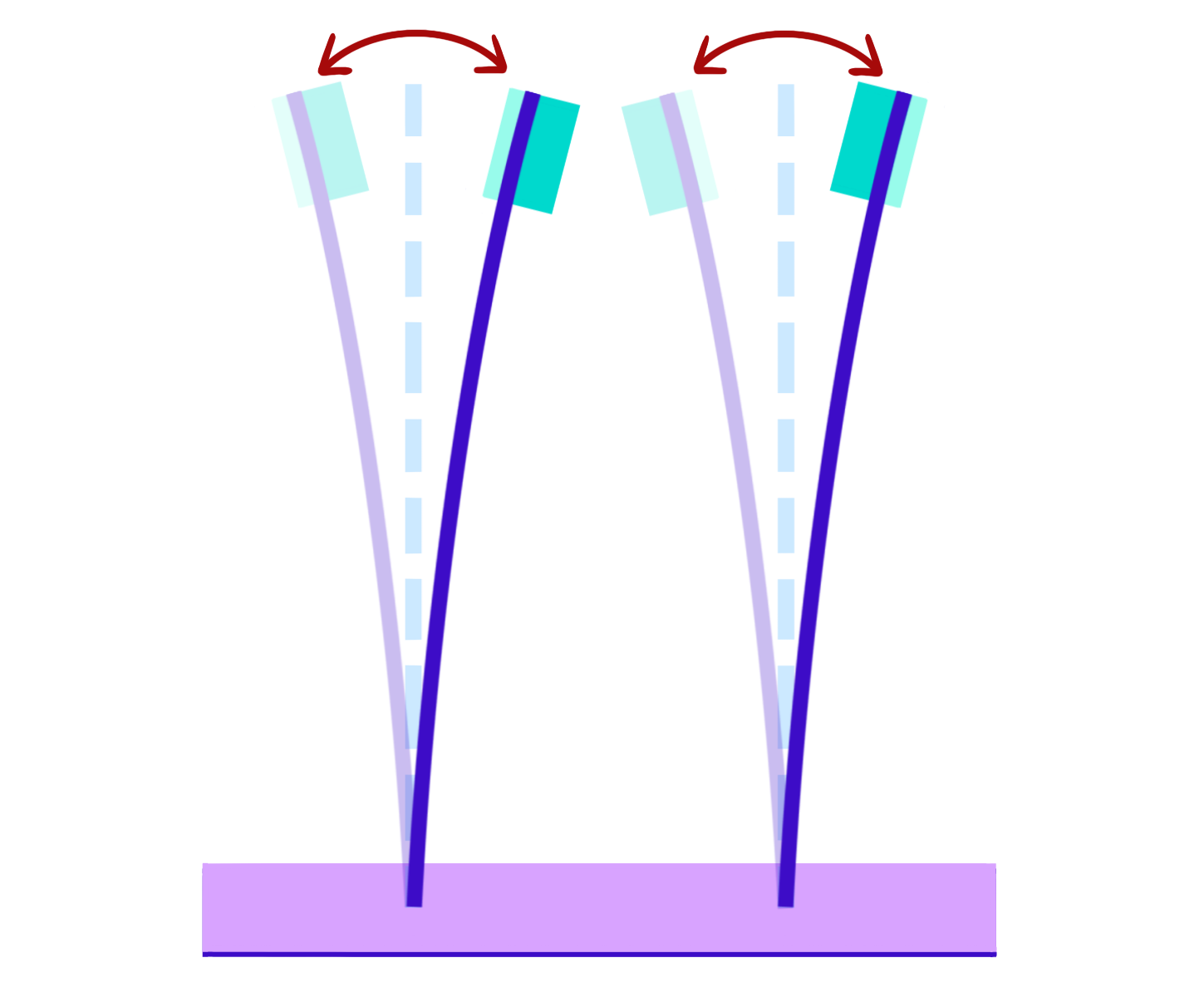}}
\end{minipage}
\caption{\textbf{Schematic diagram of the magnetic-mechanical oscillation system}
{\small (a) The overall system consists of two identical plate-shaped leaf springs, a non-magnetic base, and magnets. The leaf springs are fixed at the bottom to the non-magnetic base, and the magnets are attached to the top of the plate-shaped leaf springs.
(b) When an initial displacement is given to a single plate, it oscillates back and forth.
(c) Through magnetic coupling, the oscillation of one plate induces oscillation in the other plate as well.}}
\label{Fig.1}
\end{figure}

The main component of the system is the plate. If there is only one plate, as shown in \textbf{Fig.\ref{Fig.1(b)}}, when we give it a small initial displacement and release it, it should oscillate back and forth. However, when magnetic interactions are introduced, the motion of the two plates becomes coupled through the magnetic force. As shown in \textbf{Fig.\ref{Fig.1(c)}}, the oscillation of one plate will cause the other plate to oscillate as well.

\begin{figure}[h]
    \centering
    \includegraphics[width=0.75\textwidth]{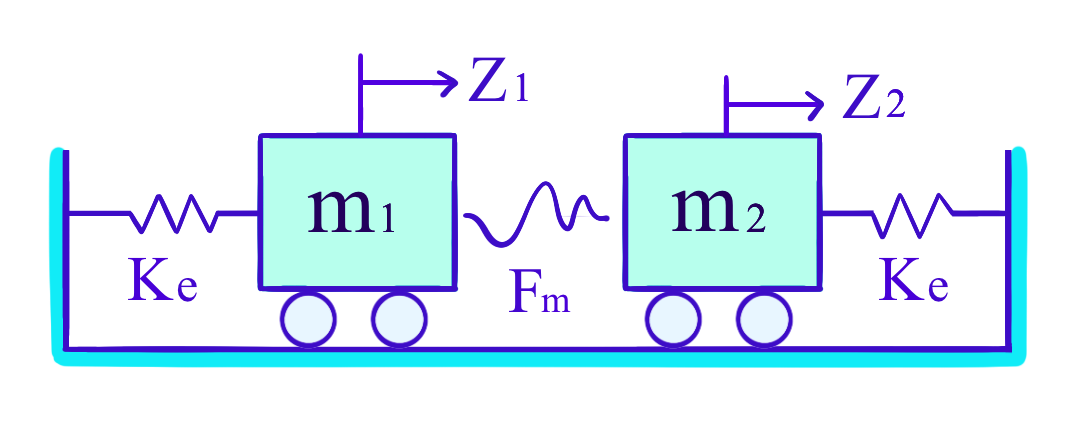}
    \caption{\textbf{We assume that the magnetic-mechanical oscillator (MMO) is similar to such system.}{\small two masses connected by springs interact through a third "spring"—the magnetic force. In this way, the motion of the two plates becomes correlated.}}
    \label{Fig.2}
\end{figure}

Therefore, we analogize the magnetic-mechanical oscillator to the system shown in \textbf{Fig.\ref{Fig.2}}: two masses connected by springs interact through a third "spring"—the magnetic force. In this way, the motion of the two plates becomes correlated.

Intuitively, it seems that the system's state can be described using two parameters $x_1$ and $x_2$, representing the displacements of 
the tips of two leaf springs. Therefore, we propose the first hypothesis: the magnetic-mechanical oscillator is a two-degree-of-freedom system.

We employ a formal set of differential equations to describe this system; let the displacements from the equilibrium positions 
of the two mass points be denoted as $x_1$ and $x_2$. Thus, a phenomenological set of coupled differential equations governing 
the magnetic-mechanical oscillator system can be formulated as follows:
\begin{equation}
  \begin{aligned}
    &\ddot{x_2} = -\displaystyle\frac{F_e(x_2)}{M_{eff}} - \displaystyle\frac{f_b(\dot{x_1},\dot{x_2})}{M_{eff}} + \displaystyle\frac{F_m(x_1 , x_2)}{M_{eff}}  
    \\ 
    &\ddot{x_1} = -\displaystyle\frac{F_e(x_1)}{M_{eff}} - \displaystyle\frac{f_b(\dot{x_1},\dot{x_2})}{M_{eff}} - \displaystyle\frac{F_m(x_1 , x_2)}{M_{eff}} 
  \label{Func.1.1}
  \end{aligned}
\end{equation}
\textbf{Table \ref{Table.1}} shows the values and meanings of the parameters mentioned.
\begin{table}[h]
  \centering
  \begin{tabular}{cc}
    \hline
    Notations & Meaning\\
    \hline
    $M_{\text{eff}}$ & Equivalent mass\\
    $F_e$ & Equivalent restoring force\\
    $f_b$ & Equivalent damping force \\
    $F_m$ & Magnetic force\\
    \hline
  \end{tabular}
  \caption{The physical quantities appearing in the differential equation and their meanings}
  \label{Table.1}
\end{table}
In the following sections, our objective is to conduct a thorough analysis of whether this model can accurately describe the 
magnetic-mechanical oscillator system. Subsequently, we will analyze the mathematical forms of these functions and establish a model 
for the dependence of these functions on basic physical quantities such as the variables $x_1$ and $x_2$, as well as material 
properties. The physical quantities that we will utilize and their respective meanings are outlined in \textbf{Table \ref{Table.2}}:

\begin{table}[h]
  \centering
  \begin{tabular}{cccc}
    \hline
    Notations & Meaning & Notations & Meanings\\
    \hline
    $a$ & Width of the plate & $\nu$ & Poisson ratio\\
    $h$ & Thickness of the plate & $\mathbf{M}$ & Magnetization\\
    $L_{all}$ & Total lenth of the plate & $D$ & Diameter of magnet\\
    $L$ & Length of the movable part & $W$ & Thickness of magnet\\
    $\rho$ & Radius of curvature & $m_0$ & Mass of magnet\\
    $\rho_e$ & Density & $m$ & Magnetic moment\\
    $Y$ & Young's modulus & $d$ & Distance between two plates \\
    \hline
  \end{tabular}
  \caption{Physical quantities we will use and their meanings}
  \label{Table.2}
\end{table}
\section{Theoretical analysis and modeling}

In this section, we divide the system into three independent parts for modular analysis, obtaining the mechanical properties of the vibration of a single plate, the interaction force between the magnets that couples the vibrations of the two plates, and the qualitative conclusions regarding the system's energy dissipation.
Basing on the above analysis, we derive the differential equations that the system satisfies and determined its dynamic properties.

\subsection{An analysis of the plate}

\subsubsection{Theoretical derivation of the plate}

We abstract the plate-shaped leaf spring as a classical spring and treat the magnet and the spring as a whole with an equivalent mass.
In this section, starting from Young's modulus and Hooke's law, we derive the expression for the spring force. Based on the quasi-static assumption, we also derive the expression for the equivalent mass \footnote[7]{Here, we only present the simplifications used in the theoretical analysis, along with the general idea and a rough outline of the derivation process. For the detailed derivation, see \textbf{Appendix A}}.

\begin{figure}[h]
\centering
\begin{minipage}[t]{0.48\textwidth}
\centering
\includegraphics[width=1\textwidth]{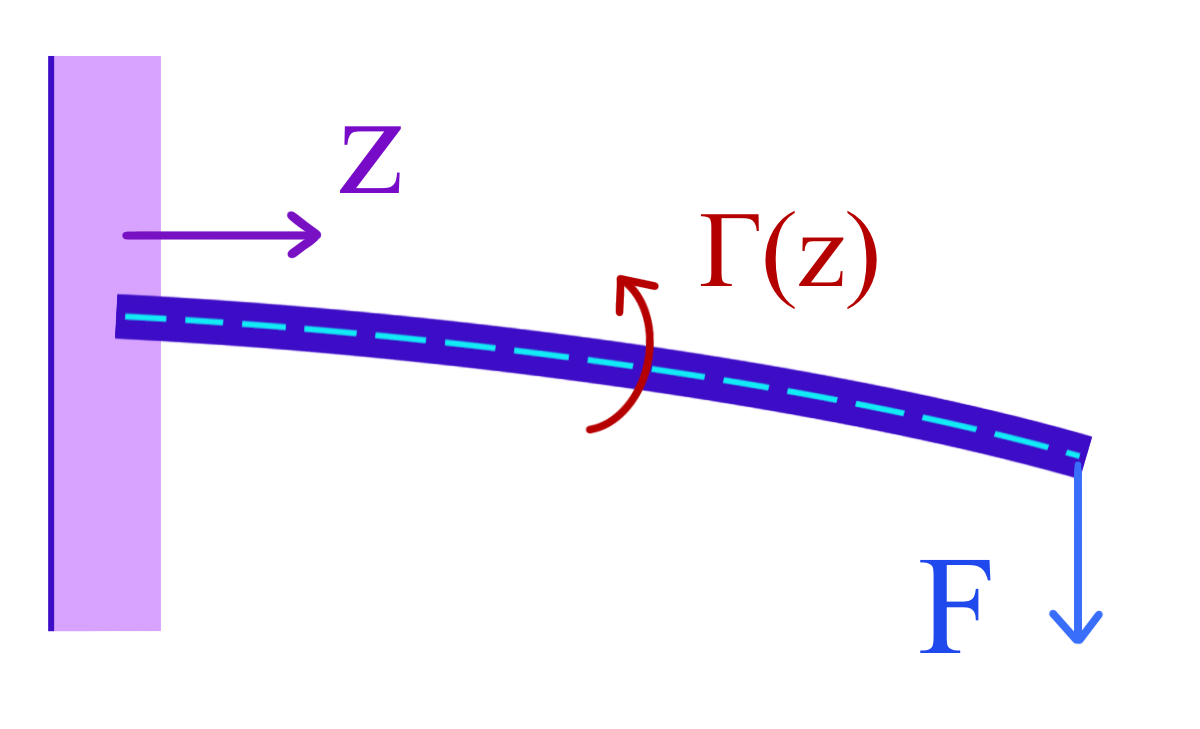}
\caption{\textbf{Torque balance diagram of a curved plate with small deflection.} T{\small he moment $\Gamma$ transmitted due to bending at the cross-section, the stress $F$ applied, and the reference position coordinate $Z$.}}
\label{Fig.3}
\end{minipage}
\hspace{9pt}
\begin{minipage}[t]{0.48\textwidth}
\centering
\includegraphics[width=1\textwidth]{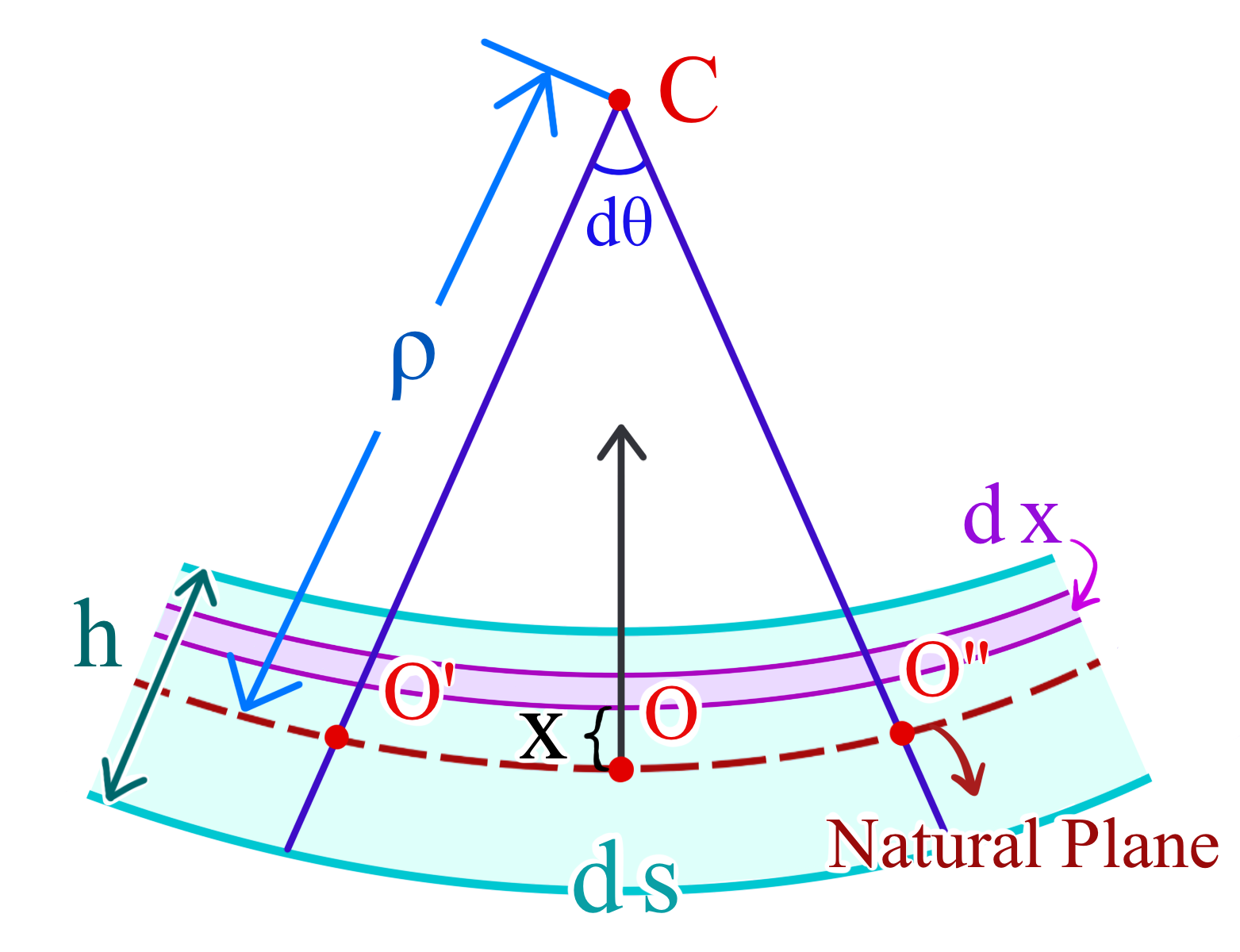}
\caption{\textbf{Details in position z.} {\small From this, we can geometrically analyze and obtain the required physical quantities at that location. At this point, we are considering a plate undergoing small deflection bending.}}
\label{Fig.4}
\end{minipage}
\end{figure}

Based on laboratory observations and its favorable geometric properties, we consider a small deflection bending of the plate as shown in \textbf{Fig.\ref{Fig.3}}, while \textbf{Fig.\ref{Fig.4}} illustrates the details at the cross-section of the plate under small deflection bending, which is the primary focus of our analysis.

The displacement of the spring's end under load is given by \textbf{$y_{end} = y|_{z=L} = \frac{4 FL^3}{Yah^3}$}. If we treat it as the deformation produced by the load $F$ applied to the equivalent spring, we can then obtain: the force applied by the plate is a linear restoring force proportional to the end displacement: $F_e = -K_e x$ (the negative sign indicates that the restoring force is in the opposite direction to the deformation), thus leading to the expression for the equivalent stiffness coefficient $K_e$:

\begin{equation}
  \begin{aligned}
    &K_e = \displaystyle\frac{F}{y_{end}} = \displaystyle\frac{Yah^3}{4L^3} 
  \end{aligned}
\end{equation}

However, we ignore the effect of strain in one direction on strain in the other direction, which is often described by Poisson's ratio $\nu$. Luckily, the nature of the linear restoring force does not change, but the expression for the equivalent stiffness coefficient $K_e$ needs to be modified\footnote[7]{for more details please refer to reference \cite{item13}}:

\begin{equation}
  \begin{aligned}
    &K_e = \displaystyle\frac{Yah^3}{4(1-\nu^2)L^3} 
  \end{aligned}
\end{equation}

Now, we calculate the system's equivalent mass from an energy perspective, but we should be careful when handling the kinetic energy of the bending plate vibration process Under the small deflection approximation, the plate's deformation pattern is nearly identical to that in the static case. Therefore, it can be assumed that the vibration equation for the plate under small amplitude satisfies the quasi-static condition: that is, the plate vibrates only in the fundamental mode.

When the end velocity is $\dot{y_{end}}$, the energy of the plate is:

\begin{equation}
  \begin{aligned}
    &y(z,t) \approx \displaystyle\frac{3y_{end}(t)}{L^3}\left( \displaystyle\frac{Lz^2}{2} - \displaystyle\frac{z^3}{6} \right)
    \\
    &E_k = \int_{0}^{L} \frac{1}{2} \rho_e a h \left(\displaystyle\frac{\partial y(z,t) }{\partial t} \right)^{2} \,dz = \frac{1}{2} M_{e} (\dot{y_{end}})^{2} 
    \\
    &M_{e} = \alpha \rho_e ahL \hspace{1cm} \alpha = 9(\frac{1}{20}+\frac{1}{252}-\frac{1}{36})
  \end{aligned}
\end{equation}

If we also consider the magnet fixed at the end of the plate, then the expression for $M_{eff}$ becomes:

\begin{equation}
  \begin{aligned}
    &M_{eff} = \alpha \rho_e ahL + m_0
  \end{aligned}
\end{equation}

In summary, we analyzed that the spring force $F_e$ of the plate is a linear restoring force: $F_e = -K_e x$ and provided the expressions for the equivalent stiffness $K_e$ and the equivalent mass $M_{eff}$.

\subsubsection{COMSOL simulation verification of the equivalent spring theory}

Based on experimental parameters, we created a model of the blade spring in the "Solid Mechanics" module of COMSOL Multiphysics. We set the two sides at the bottom of the plate, each with a length of $1.5cm$ and a width of $a$, as fixed constraints. A load $F$ was applied at the top, and we used a steady-state solver to calculate the displacement at the top under the load $F$, comparing it with the theoretical value.

We considered the movable part length of the plate $L$ and the load $F$ as two independent variables and used the "parametric sweep" function to solve our model:

\textbf{A.} Set $F$ to $0.15N$ and sweep the plate length $L$ in the range from $4$ to $15cm$.

\textbf{B.} Set $L$ to $10.5cm$ and sweep the load $F$ in the range from $0$ to $0.15N$.

\begin{figure}[h]
\centering
\begin{minipage}[t]{0.48\textwidth}
\centering
\includegraphics[width=1\textwidth]{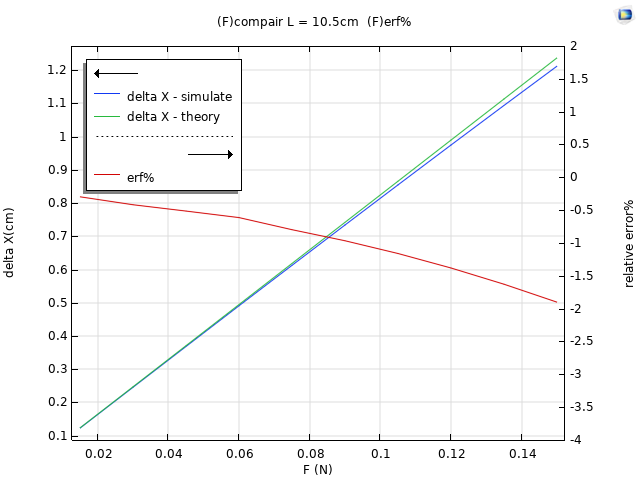}
\caption{\textbf{Constant load and sweep length.} {\small Corresponding to Setting $F$ to $0.15N$ and sweep the plate length $L_{all}$ in the range from $4$ to $15cm$.}}
\label{Fig.5}
\end{minipage}
\hspace{9pt}
\begin{minipage}[t]{0.48\textwidth}
\centering
\includegraphics[width=1\textwidth]{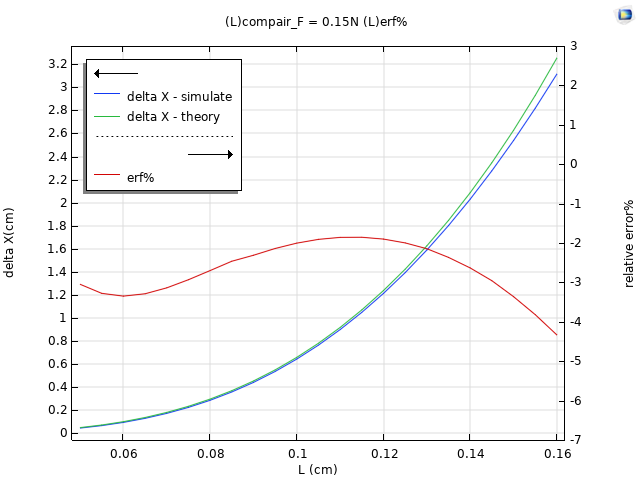}
\caption{\textbf{Constant lenth and sweep the load.}  {\small Corresponding to Set $L$ to $10.5cm$ and sweep the load $F$ in the range from $0$ to $0.15N$.}}
\label{Fig.6}
\end{minipage}
\end{figure}

We plotted the results of these two parametric sweeps and the relative error curves, as shown in \textbf{Fig.\ref{Fig.5}} and \textbf{Fig.\ref{Fig.6}}.

Within the parameter ranges considered for the two simulation sets, the theoretical curve closely aligns with the simulated curve, and the absolute value of the relative error is less than $5\%$. It can be concluded that within this 
range, our theoretical representation of the equivalent spring is effective.

\subsection{An analysis of the magnets}

\subsubsection{Theoretical derivation of the magnets}

In this section, we will analyze the form of the interaction forces between magnets \footnote[7]{Here, we only present the simplifications used in the theoretical analysis, along with the general approach and an outline of the derivation. For the detailed derivation process, please refer to \textbf{Appendix B}}. We can approximate the magnetic mechanical oscillator system to be in an environment with $\mu = \mu_0 \hspace{0.2cm} \epsilon = \epsilon_0$, and devoid of free charge $\rho_f$ and free current $\mathbf{j_f}$ ($\sigma \approx 0$) in the surroundings.
The radiation effects of the system can also be safely neglected, so we can use the magnetic scalar potential \(\varphi_m\) to describe the magnetic field around the magnetic mechanical oscillator system: \(\mathbf{H} = -\nabla\varphi_m\).

\begin{figure}[h]
\centering
\begin{minipage}[t]{0.48\textwidth}
\centering
\includegraphics[width=1\textwidth]{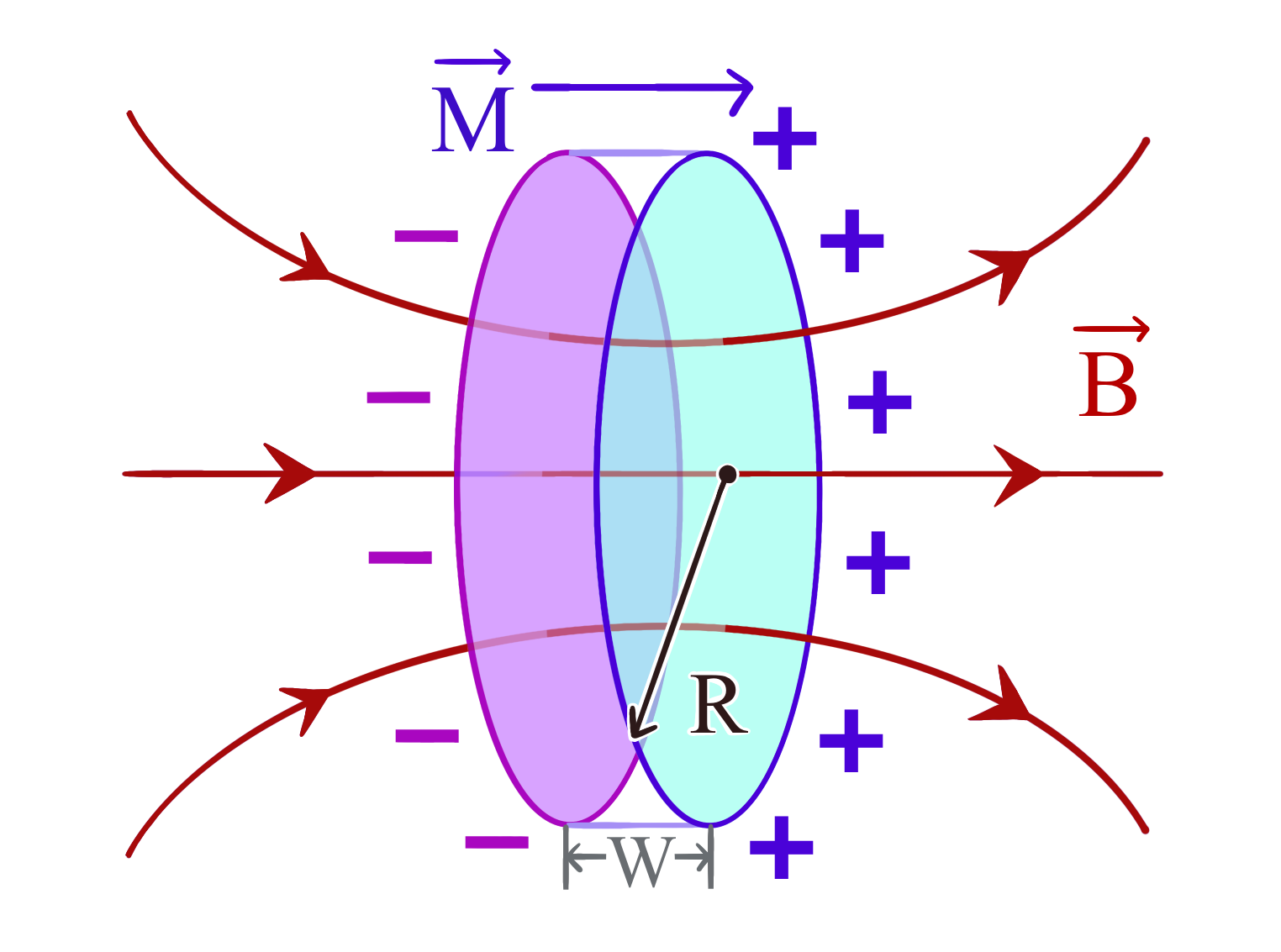}
\caption{\textbf{Diagram of a magnet.} {\small This is  a magnetic charge distributed on a geometric circle with a diameter of D,  with its axis along the z-axis.}}
\label{Fig.7}
\end{minipage}
\hspace{9pt}
\begin{minipage}[t]{0.48\textwidth}
\centering
\includegraphics[width=1\textwidth]{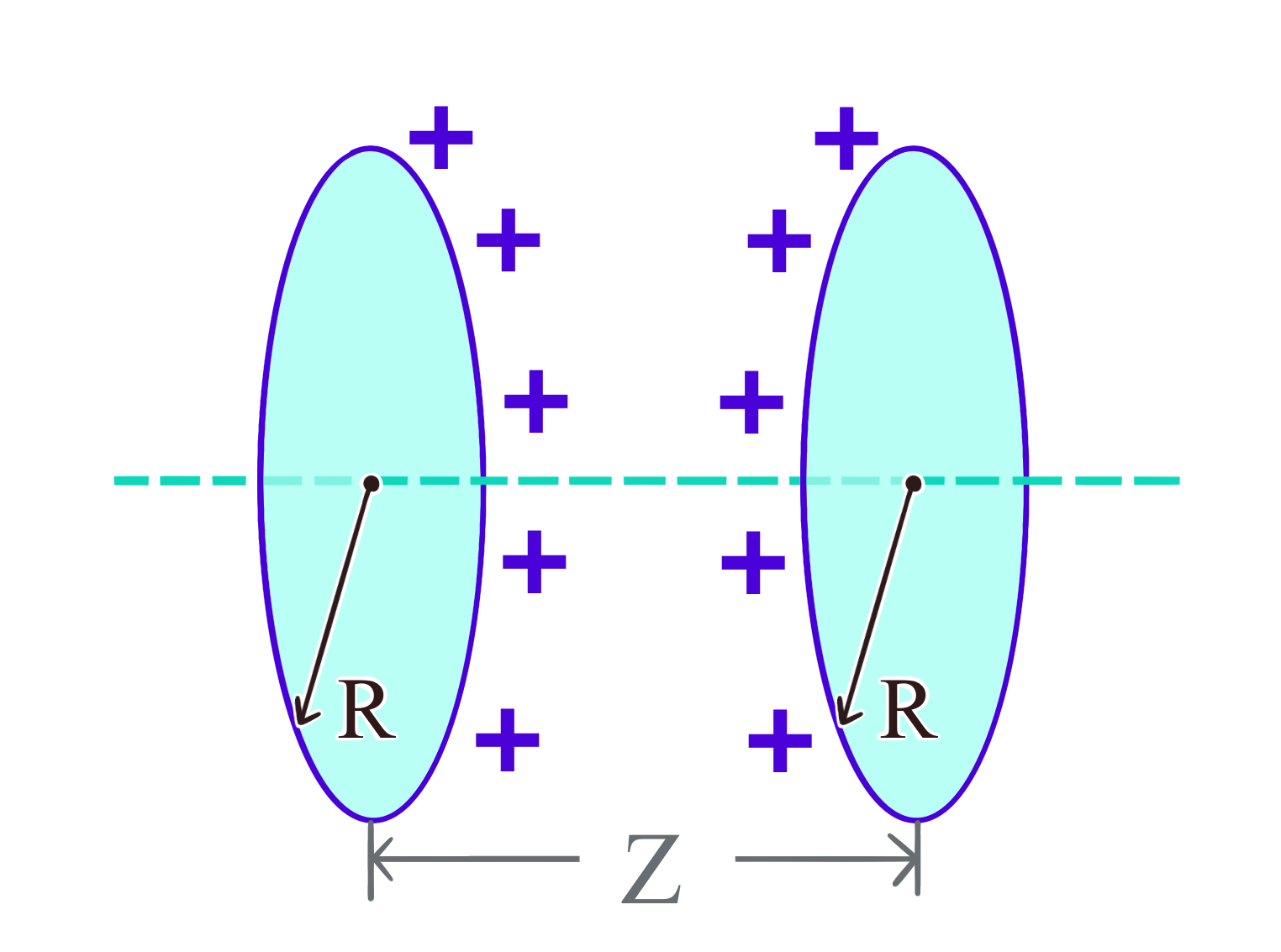}
\caption{\textbf{Two layers of magnetic charges.} {\small We consider two identical parallel and coaxially placed surface-distributed magnetic charges.}}
\label{Fig.8}
\end{minipage}
\end{figure}

In the analysis of this module, we introduce the magnetic charge theory for the magnetic field. We consider a uniformly magnetized cylindrical magnet with magnetization intensity $\mathbf{M}$ parallel to its axis as a magnetic dipole layer shown in \textbf{Fig.\ref{Fig.7}}. The surface magnetic charge density is $\sigma_m$, the diameter of the dipole layer is the same as the magnet diameter $D$, and the thickness of the magnet corresponds to the distance between the magnetic dipole layers, denoted as $W$.

Due to the small deformation of the plates, it can be approximately assumed that the two plates remain parallel to each other at all times. Therefore, the system can be simplified to two identical, parallel, and coaxially positioned surface-distributed magnetic charges, as shown in \textbf{Fig.\ref{Fig.8}}.

In summary, we obtain the equation governing the approximate force $\mathbf{F_m}$ between two magnets in the magnetic-mechanical oscillator system
\footnote[3]{If the distance between the two plates increase, the approximation presented in reference \cite{item5} can be employed, considering only the term $K_{m1}$ in the magnetic force.}.

\begin{equation}
  \begin{aligned}
    &F_m(z) = \displaystyle\frac{K_{m1}}{z^4} + \displaystyle\frac{K_{m2}}{z^6}
    \\
    &K_{m1} = \displaystyle\frac{3\mu_0 {m^2}}{2\pi} \hspace{0.5cm} K_{m2} = \displaystyle\frac{5\mu_0 m^2 (W^2 - \frac{3D^2}{4})}{4\pi}
  \end{aligned}
\end{equation}

\subsubsection{COMSOL simulation verification of the interaction force between}

We used COMSOL Multiphysics to simulate the magnetic field generated by a magnet in the air and the interaction force between two magnets. Based on this, we analyzed the validity of our proposed approximate expression for the magnetic force $(2.6)$ and compared it with the interaction force between two magnetic dipoles (considering only the $K_{m1}$ term).

\begin{figure}[h]
    \centering
    \includegraphics[width=0.7\textwidth]{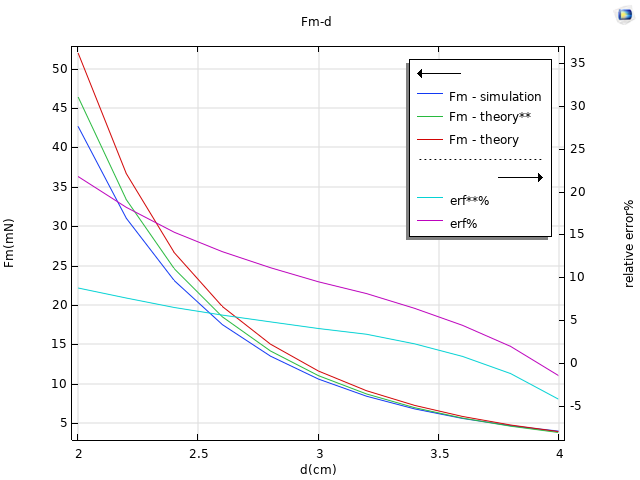}
    \caption{\textbf{Simulation of $F_m$.} {\small The graph provides the force predictions from the magnetic moment model (red line labeled theory), the high-order approximation theory we employed (green line labeled "theroy**"), as well as the simulation results (blue line labled simulation), along with the error curve. All three eventually converge.}}
    \label{Fig.9}
\end{figure}

Based on experimental parameters, we created a model of two identical cylindrical magnets in the "Magnetic Fields, No Currents" module of COMSOL. Using the built-in "Force Calculation\footnote[7]{The "Force Calculation" in COMSOL uses the Maxwell stress tensor method. To ensure the quality of the mesh, we added fillets to the magnets, but this correction was accounted for in the simulation results}" function, we calculated the interaction force $F_m$ between the two magnets. We solved this using the steady-state solver and scanned the variation of $F_m$ with $d$. The results and the error curves we obtained are shown in \textbf{Fig.\ref{Fig.9}}.

It is suggested that with an increase in distance $d$, the interaction force between the magnets rapidly decays. 
The graph provides the force predictions from the magnetic moment model (in red), the high-order approximation theory we employed (in green), as well as the simulation results (in blue), along with the error curve.

All three eventually converge, but to ensure that the magnetic force is strong enough to exhibit noticeable experimental phenomena with minimal error, the higher-order term $K_{m2}$ in the theory is necessary.

\subsection{An analysis of the loss}

The energy dissipation in the magnetic-mechanical oscillation system has several sources: air resistance, electromagnetic induction, solid deformation damping, and friction at the points where the plates are fixed. Since the plates are non-magnetic and non-conductive, the damping due to electromagnetic induction can be neglected.
Experimental measurements indicate that the system's Reynolds number$Re<1000$,so we conclude that the airflow is laminar under the experimental conditions, and air resistance is proportional to velocity.

The governing equation for solid mechanics is given by:

\begin{equation}
    M\ddot{\mathbf{u}} + C\dot{\mathbf{u}} + K\mathbf{u} = \mathbf{f}(t)
\end{equation}

where \(M\) is the mass matrix, \(C\) is the viscous damping matrix, \(K\) is the stiffness matrix, \(\mathbf{u}\) is the displacement 
vector, and \(\mathbf{f}\) is the force vector. This indicates that the damping due to solid deformation is proportional to velocity too.

Therefore, we can reasonably approximate the damping force in the system as being proportional to the velocity: $\mathbf{f_b} = -\beta \mathbf{v}$. Here, $\beta$ is very small and will not have a significant impact on the values of the two normal mode frequencies that are of most interest to us, as our experimental results later demonstrate.

\subsection{Conclusion of modeling}

Under the experimental conditions of this problem, we only consider the small oscillations of the leaf spring near its equilibrium position. At this point, the system is abstracted as two classical spring oscillators coupled through a magnetic force $F_m = \displaystyle\frac{K_{m1}}{r^4} + \displaystyle\frac{K_{m2}}{r^6}$, with the effective mass of the oscillator being $M_{eff}$, subject to an effective restoring force $F_e = -K_e x$, and a drag force proportional to the velocity.

The dependencies of each parameter are listed again as follows:

\begin{align}\left\{\begin{aligned}
    & K_e = \displaystyle\frac{Yah^3}{4(1-\nu^2)L^3} \hspace{0.6cm} M_{eff} = \alpha \rho_e ahL + m_0 \hspace{0.6cm} 
    \\
    & K_{m1} = \displaystyle\frac{3\mu_0 m^2}{2\pi} \hspace{0.6cm} K_{m2} = \displaystyle\frac{5\mu_0 m^2 (W^2 - \frac{3D^2}{4})}{4\pi}
    \\
    & \mathbf{f_b} = -\beta \mathbf{v}  \hspace{1cm} \text{not that improtant}
  \end{aligned}\right.\end{align}

By substituting our modeled results into eq. (\ref{Func.1.1}), we obtain the differential equation that describes the magnetic-mechanical oscillator:

\begin{subequations}
    \begin{equation}
        \ddot{x_2} = \displaystyle\frac{-\beta \dot{x_2}}{M_{eff}} - \displaystyle\frac{K_e x_2}{M_{eff}} + \displaystyle\frac{K_{m1}}{M_{eff}(d + x_2 - x_1)^4} + \displaystyle\frac{K_{m2}}{M_{eff}(d + x_2 - x_1)^6}
    \end{equation}
    \begin{equation}
        \ddot{x_1} = \displaystyle\frac{-\beta \dot{x_1}}{M_{eff}} - \displaystyle\frac{K_e x_1}{M_{eff}} - \displaystyle\frac{K_{m1}}{M_{eff}(d + x_2 - x_1)^4} - \displaystyle\frac{K_{m2}}{M_{eff}(d + x_2 - x_1)^6}
    \end{equation}
\end{subequations}

By performing the normal coordinate transformation$y_1 = x_1 + x_2 \hspace{0.5cm} y_2 = x_1 - x_2$,we can obtain:

\begin{subequations}
  \begin{equation}
    \ddot{y_1} = \displaystyle\frac{-\beta \dot{y_1}}{M_{eff}} - \displaystyle\frac{K_e y_1}{M_{eff}}
  \label{Func.2.10(a)}
  \end{equation}
  \begin{equation}
    \ddot{y_2} = \displaystyle\frac{-\beta \dot{y_2}}{M_{eff}} - \displaystyle\frac{K_e y_2}{M_{eff}} - \displaystyle\frac{2K_{m1}}{M_{eff}(d - y_2)^4} - \displaystyle\frac{2K_{m2}}{M_{eff}(d - y_2)^6}
  \label{Func.2.10(b)}  
  \end{equation}
\end{subequations}

These are two independent differential equations with respect to $y_1$ and $y_2$ \footnote[7]{We will only present the solution approach and some key conclusions below. For the specific solution process, please refer to \textbf{Appendix C}.}. They correspond to the symmetric and antisymmetric modes of the blade spring's motion, respectively, as illustrated in \textbf{Fig.\ref{Fig.10(a)}} and \textbf{Fig.\ref{Fig.10(b)}}:

\begin{figure}[h]
\begin{minipage}[b]{0.45\linewidth}
    \centering
    \subfloat[Symmetric mode]{\label{Fig.10(a)}\includegraphics[width=1\linewidth]{figures/sym_mode.png}}
\end{minipage} 
\medskip
\begin{minipage}[b]{0.45\linewidth}
    \centering
    \subfloat[Anti-symmetric mode]{\label{Fig.10(b)}\includegraphics[width=1\linewidth]{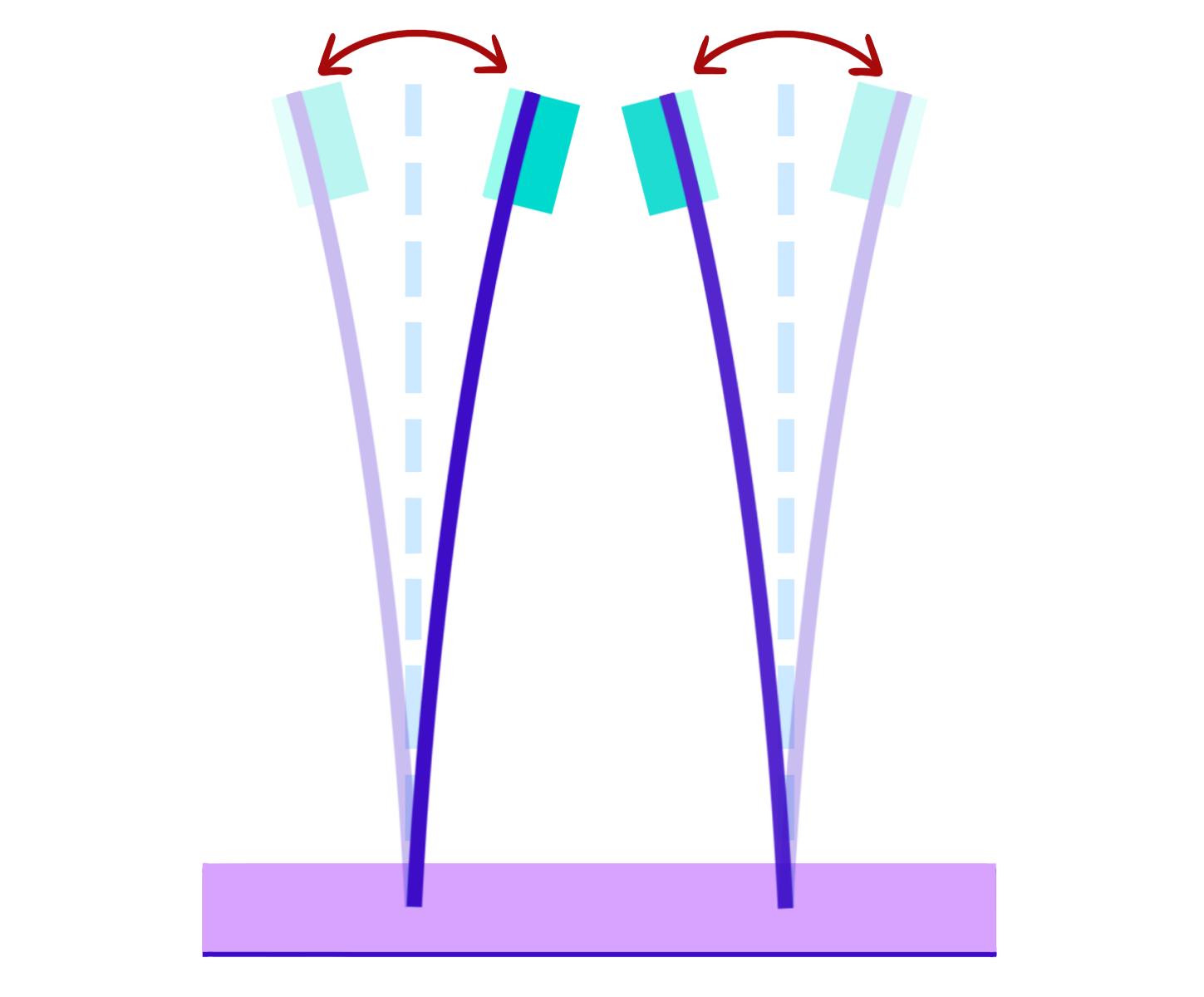}}
\end{minipage}

\caption{\textbf{The two modes of the blade spring's motion.} 
\\
{\small (a). Symmetry mode, 
in which the two leaf springs move in the same direction with identical amplitude, symmetry mode can be solved by Eq. (\ref{Func.2.10(a)})
\\
(b). Anti Symmetry mode, in which the two leaf springs move in different direction and with identical amplitude, anti symmetry mode can be solved by Eq. (\ref{Func.2.10(b)}).}}
\end{figure}

By applying the initial conditions, we can solve for $y_1(t)$ and $y_2(t)$ .Then, using the inverse transformation $x_1 = \displaystyle\frac{y_1 + y_2}{2} \hspace{0.2cm} x_2 = \displaystyle\frac{y_1 - y_2}{2}$ ,we can obtain $x_1(t)$ and $x_2(t)$.

Due to the presence of damping terms, the amplitudes of both modes will eventually approach 0. However, what is unusual is that $y_2$ will ultimately approach a finite value $y_{2e}$, where $y_{2e}$ is the solution to the following nonlinear equation:

\begin{equation}
    K_e x = \displaystyle\frac{2K_{m1}}{(d - x)^4} - \displaystyle\frac{2K_{m2}}{(d - x)^6}
\end{equation}

This is because the inherent repulsive force between the magnets causes both plates to be slightly bent outward in their equilibrium state. During the experimental data processing, we took this into account, adjusting the initial positions of the two plates during calibration as follows:

\begin{equation}
  \begin{aligned}
    & x_2 = \displaystyle\frac{y_{2e}}{2} \hspace{0.5cm} x_1 = -\displaystyle\frac{y_{2e}}{2}  
  \end{aligned}
\end{equation}

Due to the non-zeros statistical nature of the mechanical system, during the plotting process, we adjusted the initial positions to: $x_2 = \displaystyle\frac{y_{2e}}{2} \hspace{0.5cm} x_1 = -\displaystyle\frac{y_{2e}}{2}$

ext, we analyze these two equations separately:

Equation (\ref{Func.2.10(a)}) is a simple second-order differential equation with constant coefficients. Solving it yields the angular frequency of the independent vibration of one plate (with a magnet on top):

\begin{equation}
  \omega_1 = \sqrt{\displaystyle\frac{K_e}{M_{eff}} - \displaystyle\frac{\beta^2}{4M_{eff}^2}} \approx \sqrt{\displaystyle\frac{K_e}{M_{eff}}}
\end{equation}

Equation (\ref{Func.2.10(b)}) is a nonlinear differential equation. We can handle it effectively using substitution and successive approximation methods for nonlinear vibration introduced in reference \cite{item14}. At this point, we have:

\begin{align}
    \Omega &= \sqrt{\displaystyle\frac{K_e}{M_eff} + \displaystyle\frac{8K_m}{M_{eff}d^5} -\displaystyle\frac{2\alpha\theta}{\omega_{20}^2} +\displaystyle\frac{3\beta \theta^2}{\omega_{20}^{4}}}
    \\
    \omega_2 &\approx \Omega + \left[ \displaystyle\frac{3\beta}{8\Omega} - \displaystyle\frac{5\left(\alpha - \frac{3\beta\theta}{\omega_{20}^2} \right)^2}{12\Omega^3} \right]A^2
\end{align}

where some coefficients we used are listed here, they will rise from taylor series:

\begin{align}\left\{\begin{aligned}
    \alpha &= \displaystyle\frac{20K_{m1}}{M_{eff}d^6} + \displaystyle\frac{42K_{m2}}{M_{eff}d^8} \hspace{0.5cm} \\ \beta &= \displaystyle\frac{40K_{m1}}{M_{eff}d^7} + \displaystyle\frac{112K_{m2}}{M_{eff}d^9} \hspace{0.5cm} \\ \theta &= \displaystyle\frac{2K_m}{M_{eff}d^4} + \displaystyle\frac{2K_{m2}}{M_{eff}d^6}
\end{aligned}\right.\end{align}

Here, we define a dimensionless number $Z$ to characterize the relative strength of the elastic force to the magnetic force:

\begin{equation}
  Z := \displaystyle\frac{K_e d^5}{K_{m1}}
\end{equation}

$Z$ can be used to determine under what conditions our theory is valid.

\section{Experimental materials and methods}

\subsection{Experimental materials}

1. Plate-related Parameters

We purchased a type of fiberglass material that is non-magnetic, nearly insulating, and exhibits good elasticity, making it an ideal material for our experiments. The mechanical parameters were provided by the manufacturer, and we fabricated several plates of identical specifications from this material. The relevant parameters are shown in \textbf{Table \ref{Table.3}}:

\begin{table}[h]
  \centering
  \begin{tabular}{cccc}
    \hline
    Notations & Value & Geometry & Value\\
    \hline
      $\rho_e$ & $2.25g\cdot cm^{-3}$ & $h$ & 0.500 mm\\
      $Y$ & 71.9 GPa & $a$ & 6.00 mm\\
      $\nu$ & 0.2 & $L_{all}$ & 16.00 cm\\ 
    \hline
  \end{tabular}
  \caption{Physical quantities and their values related to the plate}
  \label{Table.3}
\end{table}

\begin{figure}[h]
\centering
\begin{minipage}[t]{0.48\textwidth}
\centering
\includegraphics[width=1\textwidth]{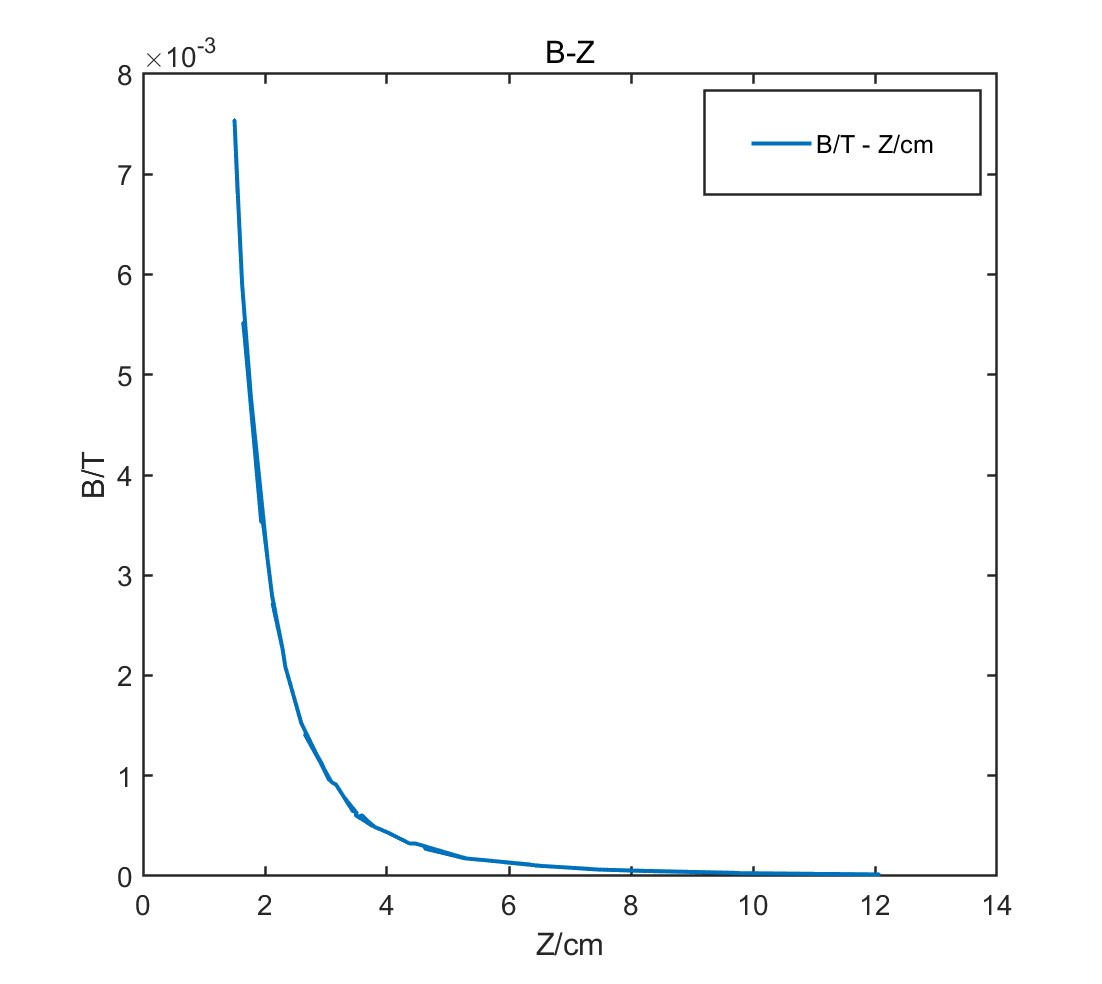}
\caption{\textbf{$B_z$ data.} {\small The raw data of measuring the component of the magnetic field $B_z$ along the axis of the cylindrical magnet.}}
\label{Fig.11}
\end{minipage}
\hspace{9pt}
\begin{minipage}[t]{0.48\textwidth}
\centering
\includegraphics[width=1\textwidth]{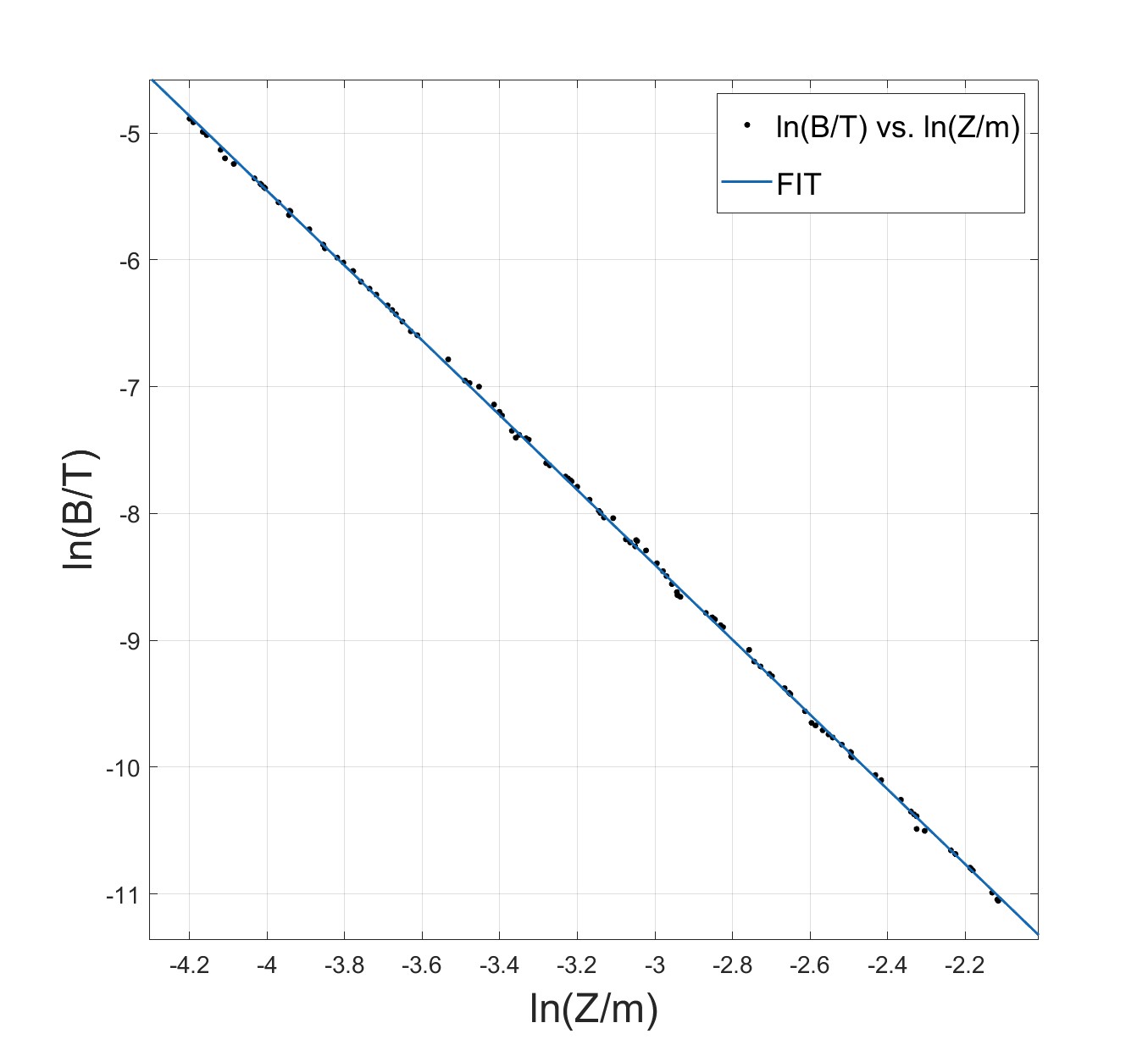}
\caption{\textbf{Curv fitting.} {\small The result of using the MATLAB curve fitting tool to fit the data corresponding to the experimental region after plotting the double logarithmic. We fit those data points with $y = p1*x + p2$ curve, where p1 $\approx$ -2.95, p2 $\approx$ -17.37 with $R^2$ $\approx$ 0.9997}}
\label{Fig.12}
\end{minipage}
\end{figure}

2. Magnet-related Parameters:

We use displacement sensors and 3D magnetic field sensors to measure the magnetic field in the space surrounding the magnets. For simplicity, we measure the component of the magnetic field $B_z$ along the axis of the cylindrical magnet. The raw data is plotted in \textbf{Fig.\ref{Fig.11}}.

The magnetic field formula generated by a magnetic moment is given by:

\begin{equation}
    \mathbf{B} = \frac{\mu_0 m}{4\pi r^3}\left(2\cos{\theta} \hat{\mathbf{r}} + \sin{\theta}\hat{\mathbf{\theta}}\right)
\end{equation}

We obtain that the expected value of the magnetic field component along the axis parallel to the magnet's axis should satisfy:

\begin{equation}
  \begin{aligned}
  &\ln{B_z} = \ln{\displaystyle\frac{\mu_0 m}{2 \pi}} - 3 \ln{Z}
  \end{aligned}
\end{equation}

We create a double logarithmic plot of $log(B_z) - log(z)$ and use the MATLAB curve fitting tool to fit the data corresponding to the experimental region. The results obtained are shown in \textbf{Fig.\ref{Fig.12}}.

The fitted values are as follows:

\begin{equation}
 \begin{aligned}
  & ln(B_z/T) = p_1 ln(z/m) + p_2 
  \\
  & p_1 = -2.95 \pm 0.04 \hspace{0.5cm} p_2 = -17.37 \pm 0.15 \hspace{0.5cm} R^2 = 0.9997
 \end{aligned}
\end{equation}

From this, we can derive the expression for calculating $m$:

\begin{equation}
 m = \displaystyle\frac{2 \pi}{\mu_0} e^{p_2}
\end{equation}

Based on our measurements and calculations, we obtained some parameters related to the magnets, as shown in \textbf{Table \ref{Table.4}} (parameters for two magnets):

\begin{table}
  \centering
  \begin{tabular}{cc}
    \hline
    Notations & Value\\
    \hline
      $D$ & 9.40 mm\\
      $W$ & 2.000 mm\\
      $m_0$ & 2.080 g\\ 
      $m$ & $0.1424  A\cdot m^{2} $\\
      $K_{m1}$ & $1.217 \times 10^{-8} N\cdot m^4$\\
      $K_{m2}$ & $-6.314 \times 10^{-13} N\cdot m^6$\\
    \hline
  \end{tabular}
  \caption{Physical quantities and their values related to the magenet}
  \label{Table.4}
\end{table}

\subsection{Experimental methods}

\begin{figure}[h]
\begin{minipage}[b]{0.6\linewidth}
    \centering
    \subfloat[Full sight of the instruments]{\label{Fig.13(a)}\includegraphics[width=0.93\linewidth]{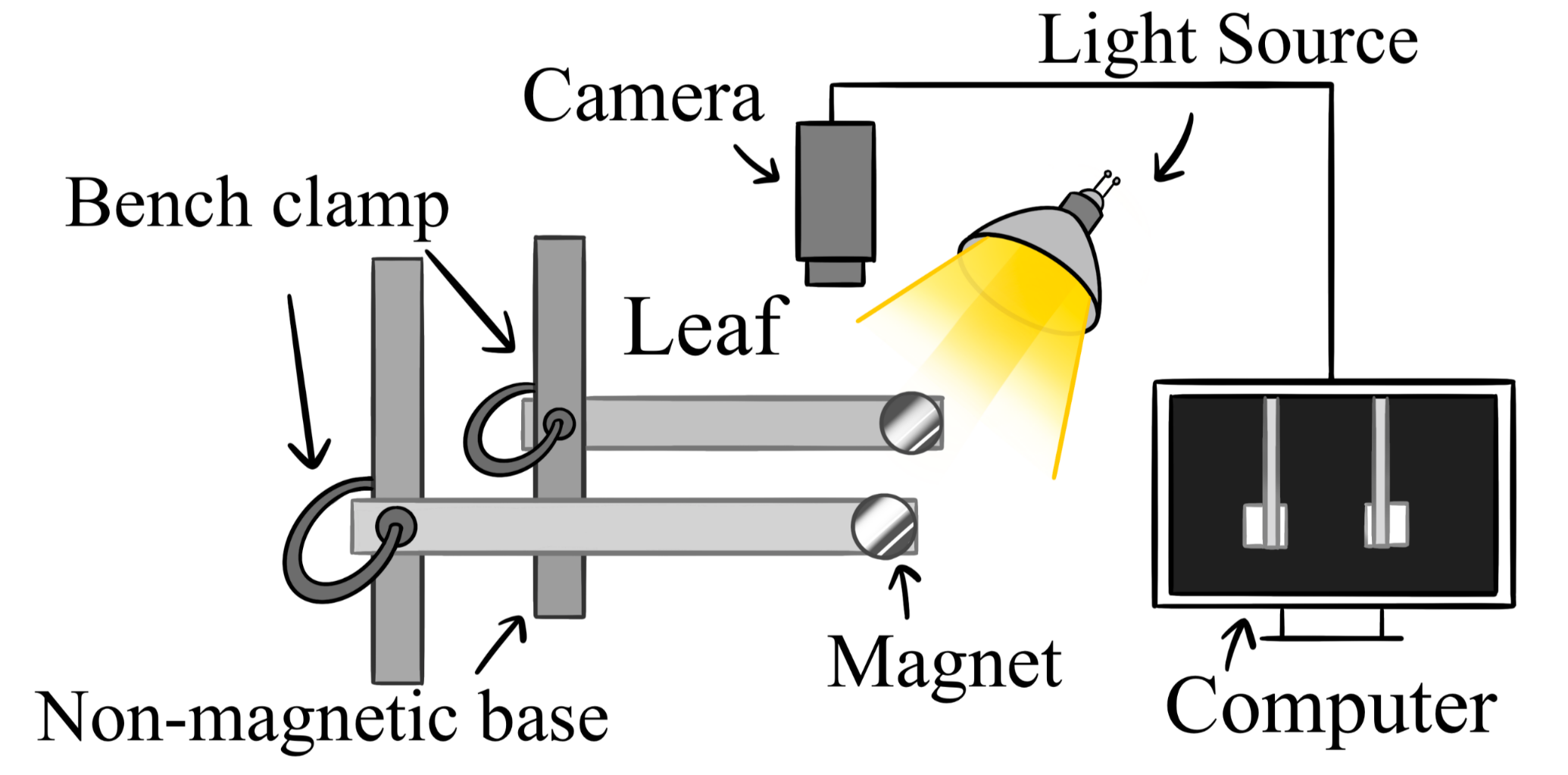}}

    \subfloat[Fixing way, Vertical view and Side view]{\label{Fig.13(b)}\includegraphics[width=0.93\linewidth]{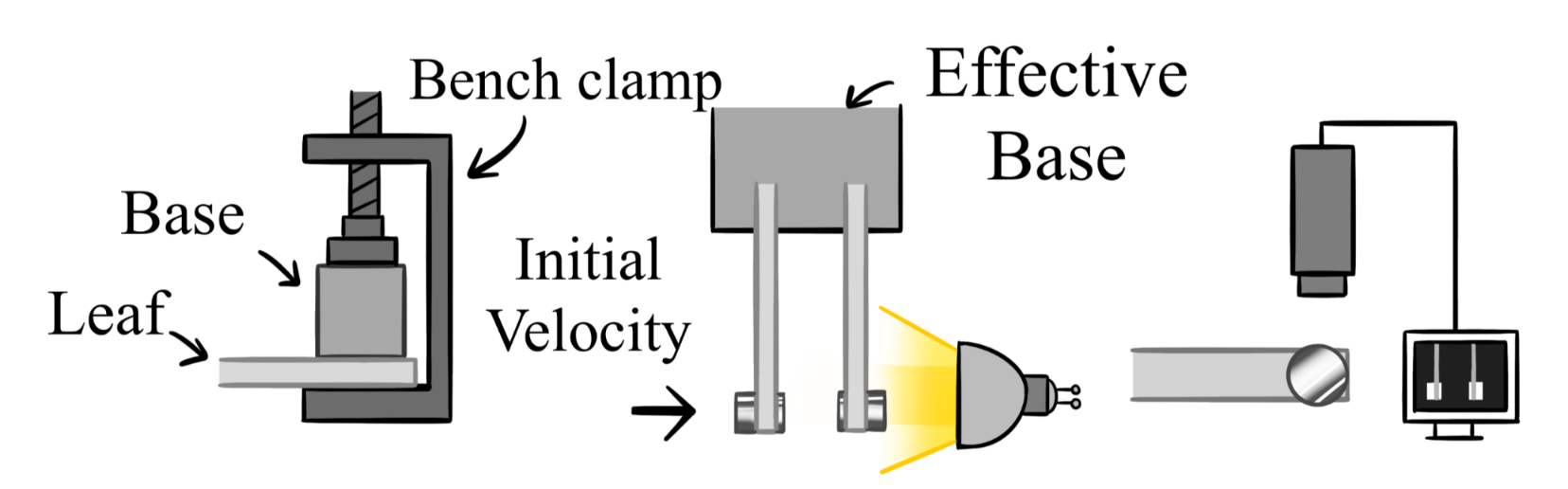}}
\end{minipage} 
\medskip
\begin{minipage}[b]{0.4\linewidth}
    \centering
    \subfloat[Starting oscillation]{\label{Fig.13(c)}\includegraphics[width=0.93\linewidth]{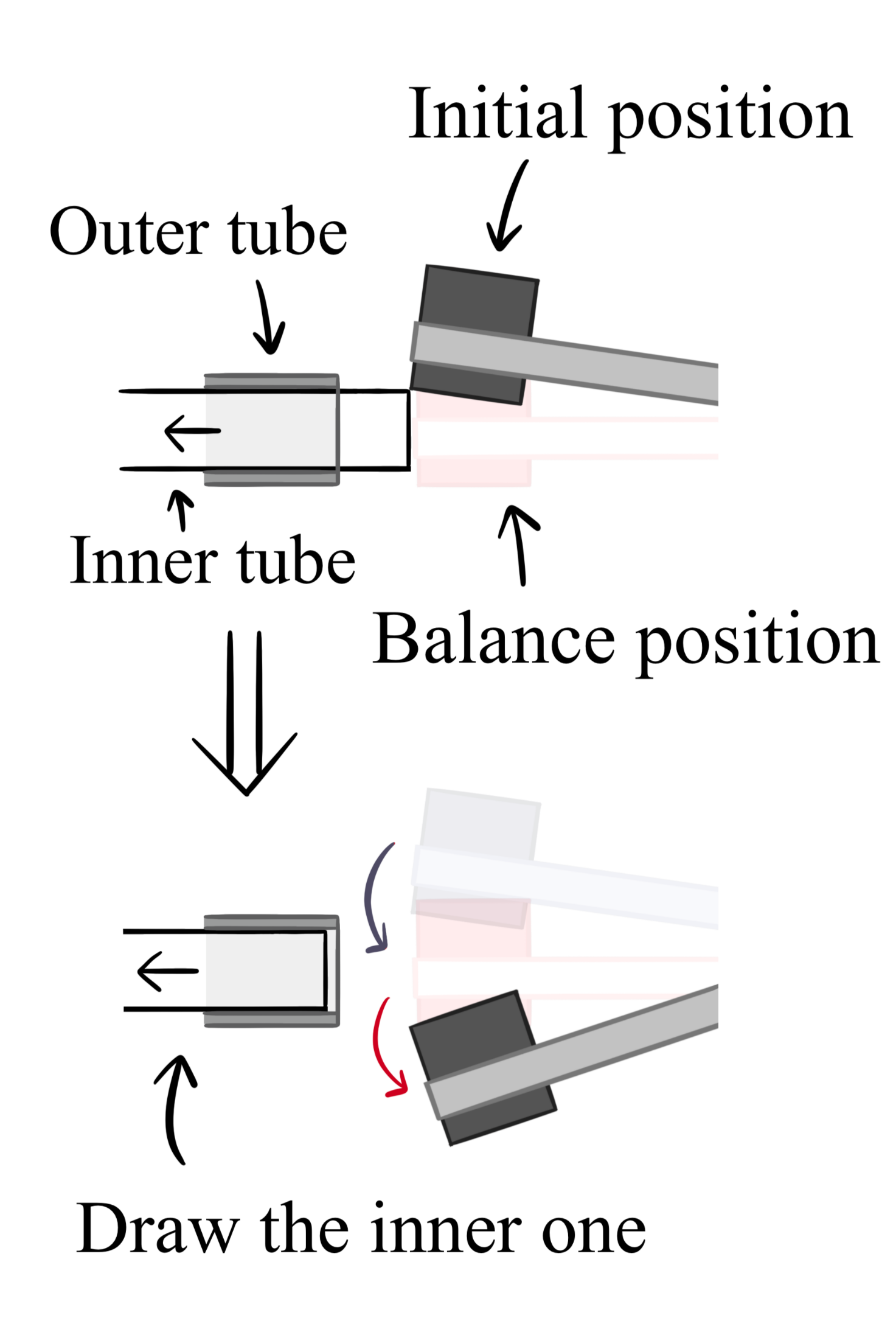}}
\end{minipage}

\caption{\textbf{Schematic diagram of the experimental setup.} {\small (a) is the panoramic view of the experiment, showing the overall layout of the experimental setup. The fixture clamps the leaf springs horizontally onto a non-magnetic base, with magnets attached to the top of the springs. A high-power light source illuminates the magnets, and a high-speed camera connected to a computer captures vertical images.
(b) provides details on how the experimental setup is fixed, including side and top views of the main experiment.
(c) illustrates how we induce vibrations in the system. By pulling out the pin as indicated by the arrows, the leaf springs are precisely activated.}}
\label{Fig.13}
\end{figure}

\textbf{Figure \ref{Fig.13}} shows the design of our experimental setup. As depicted in the overview in \textbf{Fig.\ref{Fig.13(a)}}, we symmetrically secured the fiberglass plates horizontally using a vise, which was pre-calibrated with a level to ensure that the setup was perfectly horizontal\footnote[7]{This horizontal alignment was pre-calibrated using a level, which helps prevent the influence of gravity on the vibration of the plates.}. The neodymium magnets were symmetrically attached to the top ends of the plates with their poles arranged to repel each other, as detailed in \textbf{Fig.\ref{Fig.13(b)}}.

For the excitation process of this system, we designed and fabricated a release device, as shown in \textbf{Fig.\ref{Fig.13(c)}}. By pulling out a pin, we can release the leaf spring without any initial velocity, controlling the initial displacement to be $0.50 \pm 0.05 cm$, which is more reliable than manually triggering the spring.

Our formal experiment was divided into two groups:

\textbf{A:} The distance d was controlled at 3.00 cm, and the length L of the moving part of the plate was changed at regular intervals.

\textbf{B:} The length L was controlled at 10.50 cm, and the distance d was changed at regular intervals.

After releasing one of the plates using the displacement device, the magnetic-mechanical oscillator begins to oscillate, and it can be directly observed that both plates oscillate back and forth with alternating amplitude.

We used an industrial camera, the WP-GUT130 from HUAGUDONGLI, set at 800 frames per second for vertical downward shooting to record the displacement of the tops of the two plates over time. During filming, we used a high-power light source to illuminate the magnets and laid black velvet underneath the system. The bright spots from the light reflected by the magnets were used to track the positions of the magnets, aiding in the accuracy of the tracking. Details of the setup are shown in \textbf{Fig.\ref{Fig.13(a)}} and \textbf{Fig.\ref{Fig.13(b)}}.

For the recorded video data, we first used the "Perspective Filter" feature in the Tracker software to eliminate any residual parallax. We then tracked the position of the bright spots, obtaining the data of the magnets' displacement over time.

\subsection{Results of our Experiments}

\begin{figure}[h]
\begin{minipage}[b]{0.48\linewidth}
    \centering
    \subfloat[Energy transfer]{\label{Fig.14(a)}\includegraphics[width=1\linewidth]{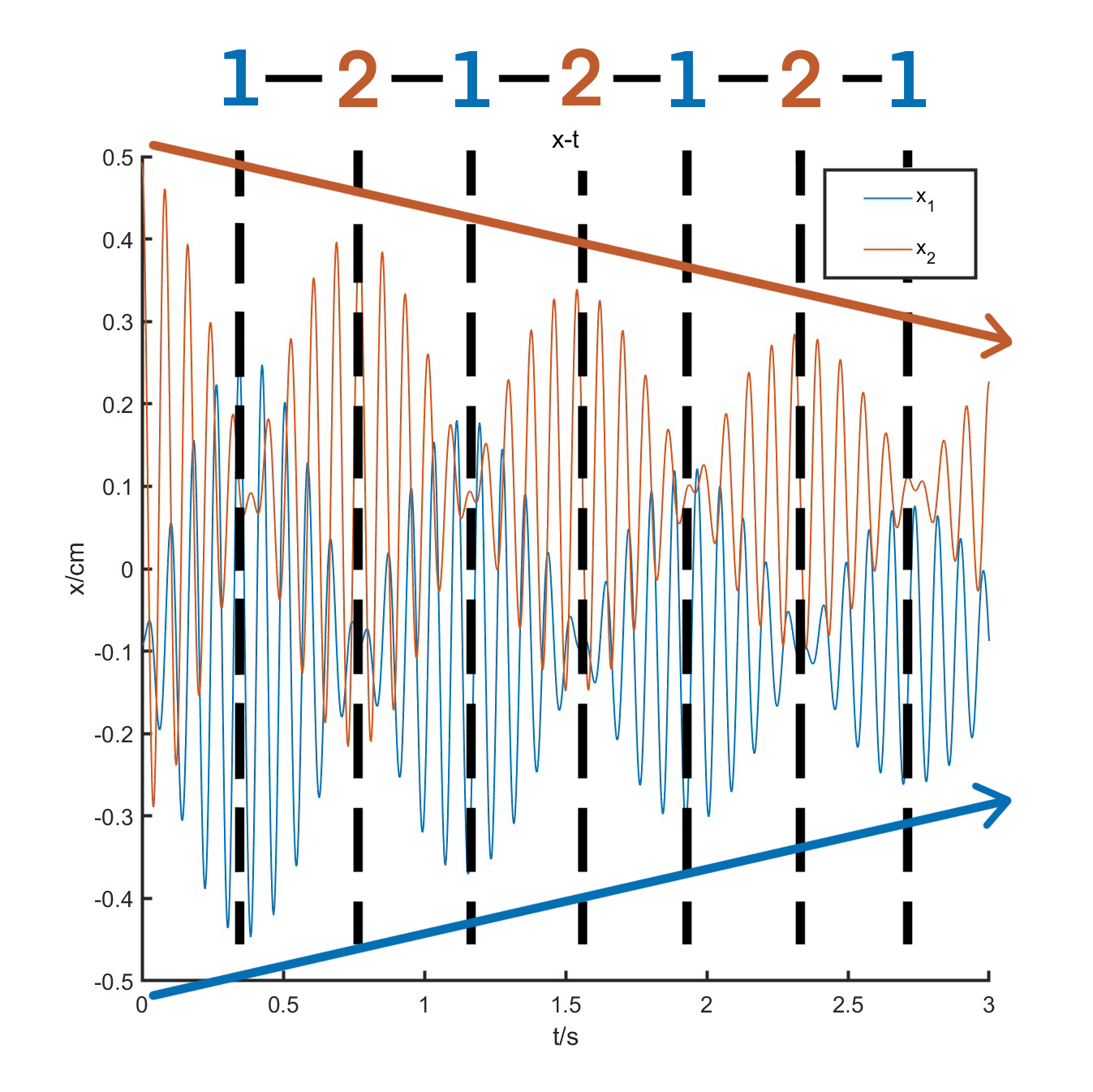}}
\end{minipage} 
\medskip
\begin{minipage}[b]{0.48\linewidth}
    \centering
    \subfloat[Energy loss]{\label{Fig.14(b)}\includegraphics[width=1\linewidth]{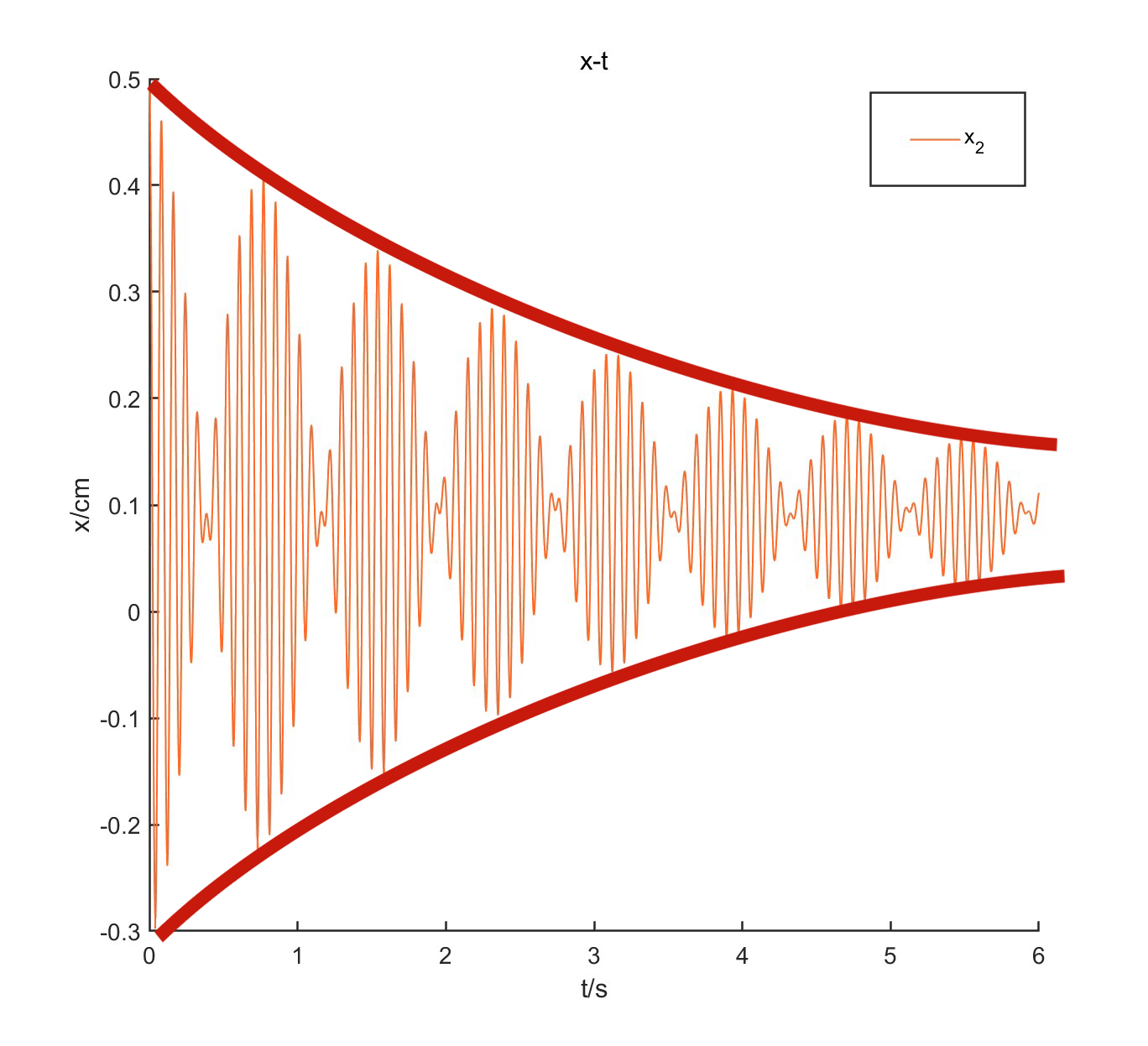}}
\end{minipage}

\caption{\textbf{The two modes of the blade spring's motion.} {\small In (a), on a smaller time scale, the overlapping data of the two plates from the same set of measurements clearly demonstrate the alternating exchange of energy. In (b), on a larger time scale, the energy dissipation process becomes evident. At this point, it is also possible to calculate the dissipation coefficient by plotting the outer envelope.}}
\label{Fig.14}
\end{figure}

\textbf{Figure\ref{Fig.14}} shows a set of displacement-time graphs we obtained. \textbf{Figure\ref{Fig.14(b)}} shows that on a smaller time scale, the amplitudes of the two leaf springs exhibit a "beat" characteristic: each spring oscillates with its own period, while the amplitude of the oscillation is modulated by another periodically varying function. The amplitude of each spring alternates in size because the magnetic coupling between the two causes energy to transfer between them. 

We can observe that due to the presence of dissipation in the system, the amplitudes of both springs gradually decrease over time. On a larger time scale, as shown in \textbf{Fig.\ref{Fig.14(b)}}, the dissipation effect becomes more pronounced, and we can determine the dissipation coefficient by plotting the envelope of the amplitude decay.

\begin{figure}[h]
\begin{minipage}[b]{0.5\linewidth}
    \centering
    \subfloat[$A,\omega_1$]{\label{Fig.15(a)}\includegraphics[width=0.95\linewidth]{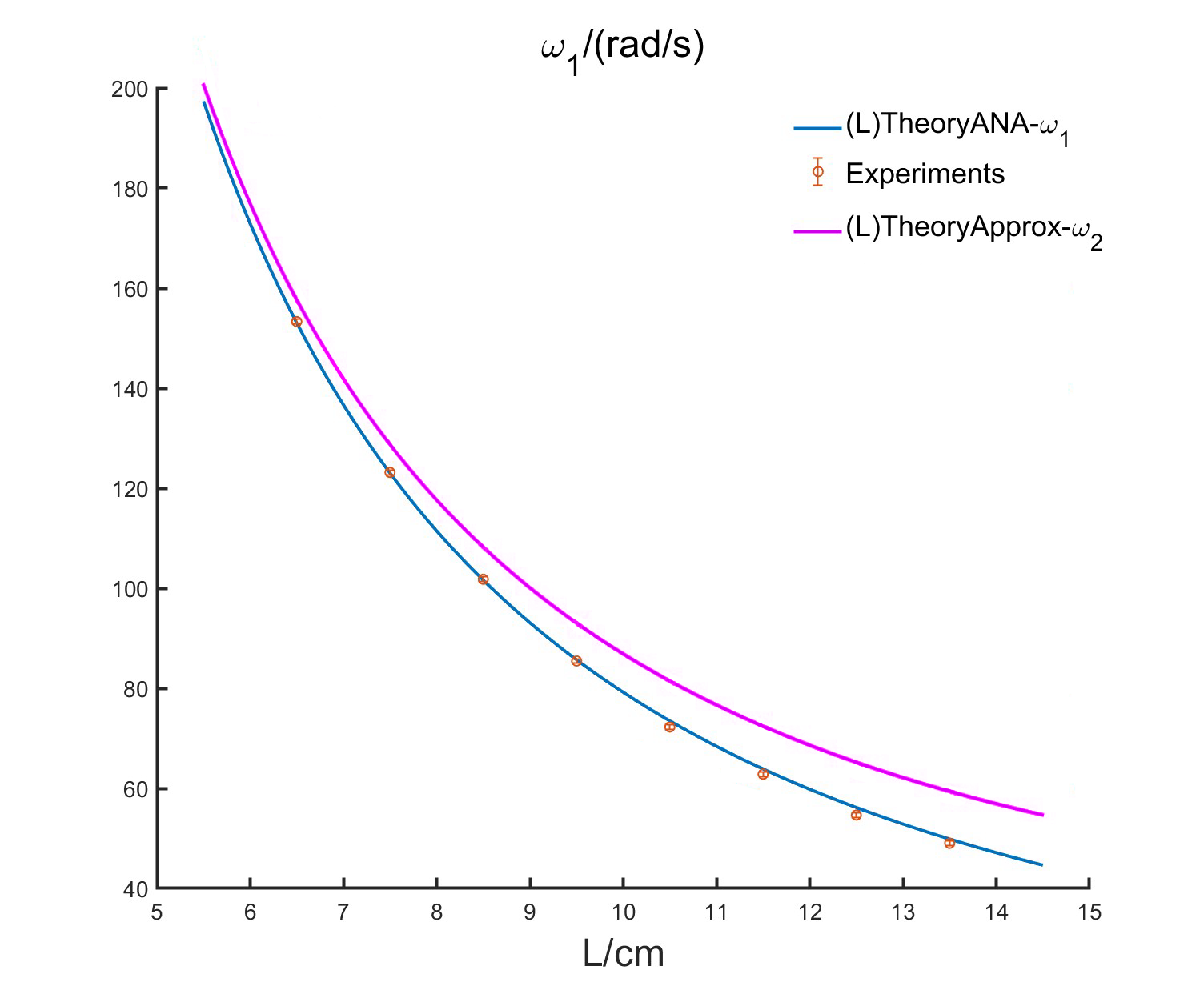}}

    \subfloat[$A,\omega_2$]{\label{Fig.15(b)}\includegraphics[width=0.95\linewidth]{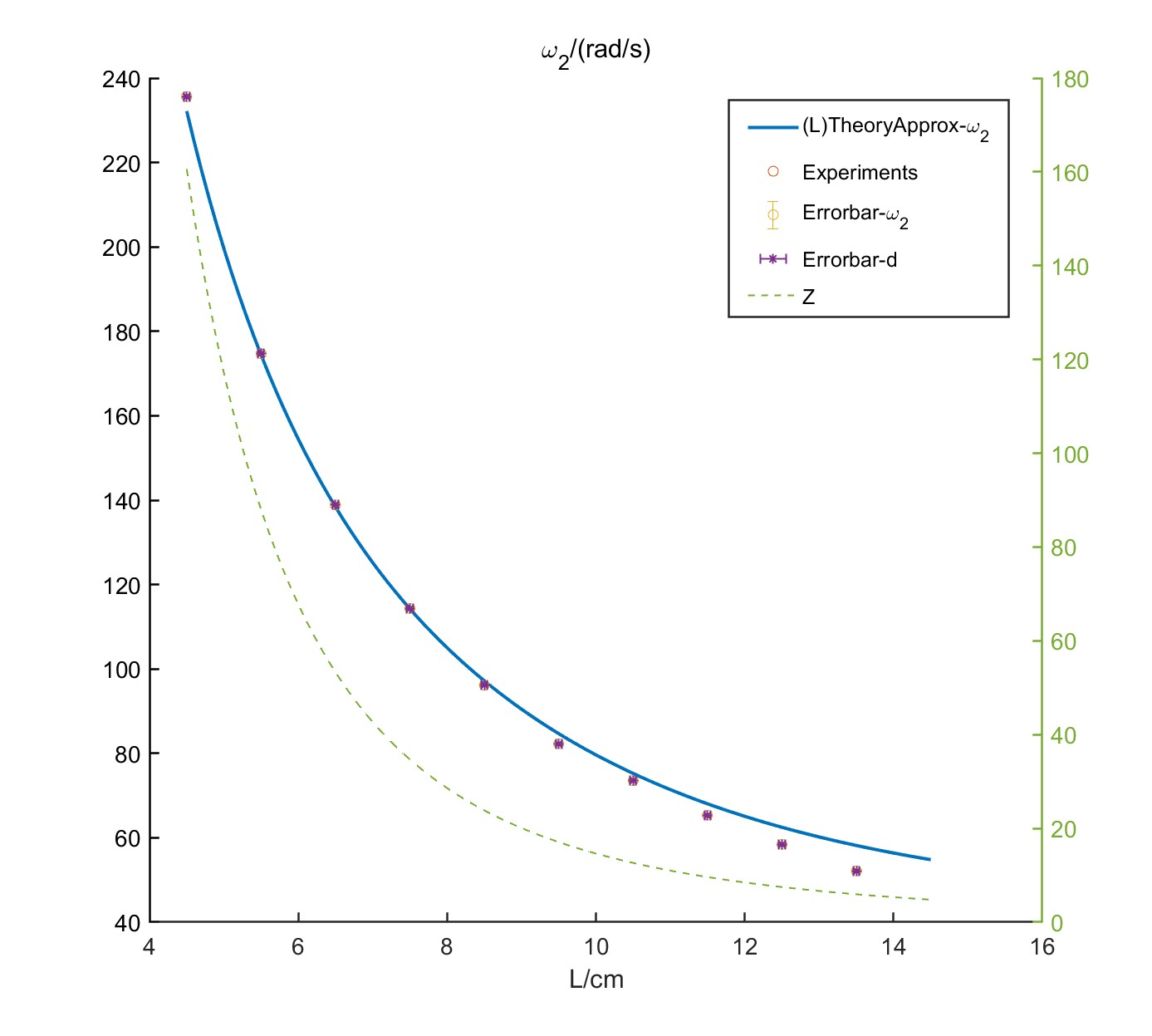}}
\end{minipage} 
\medskip
\begin{minipage}[b]{0.5\linewidth}
    \centering
    \subfloat[$B,\omega_1$]{\label{Fig.15(c)}\includegraphics[width=0.95\linewidth]{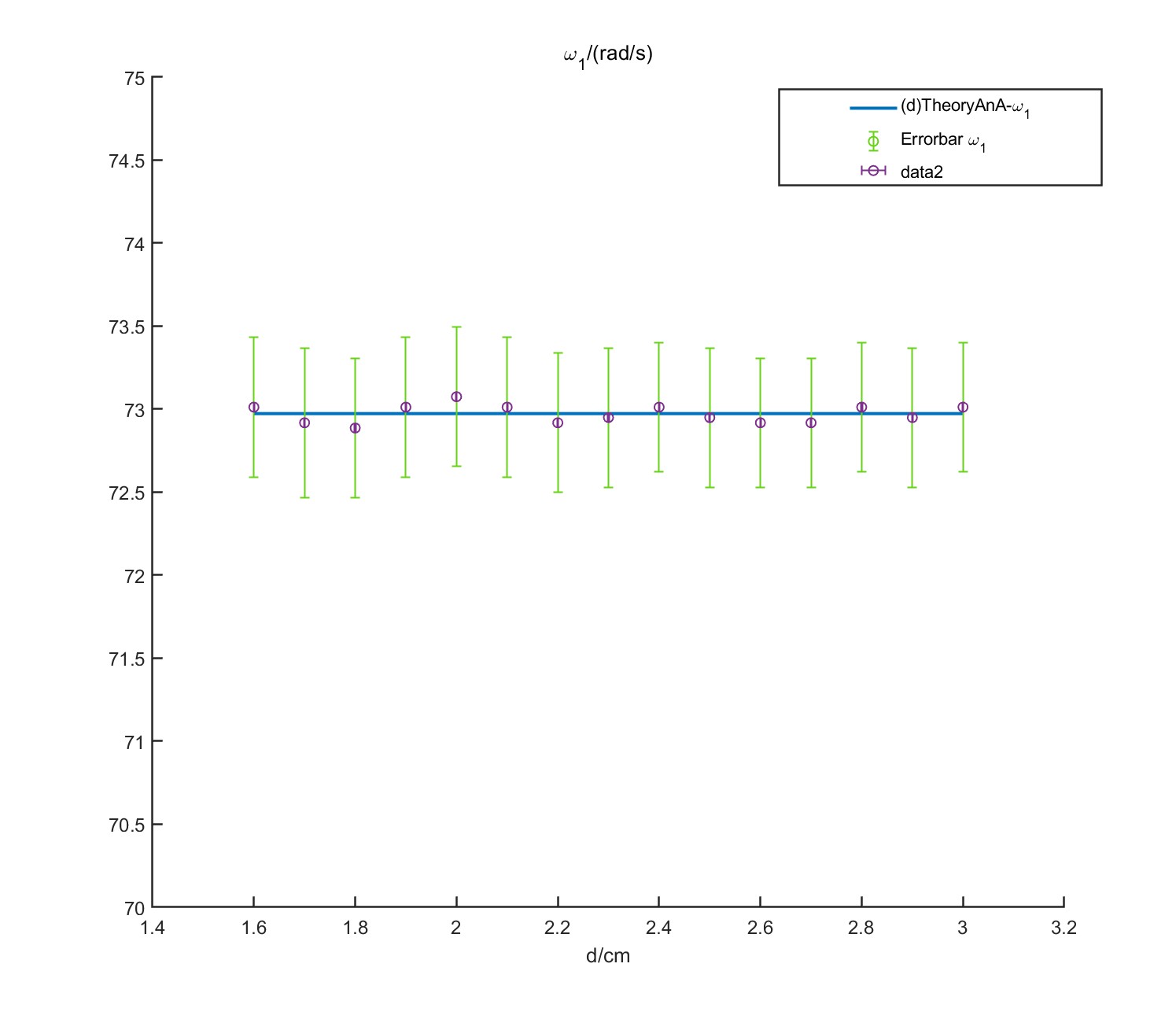}}

    \subfloat[$B,\omega_2$]{\label{Fig.15(d)}\includegraphics[width=0.95\linewidth]{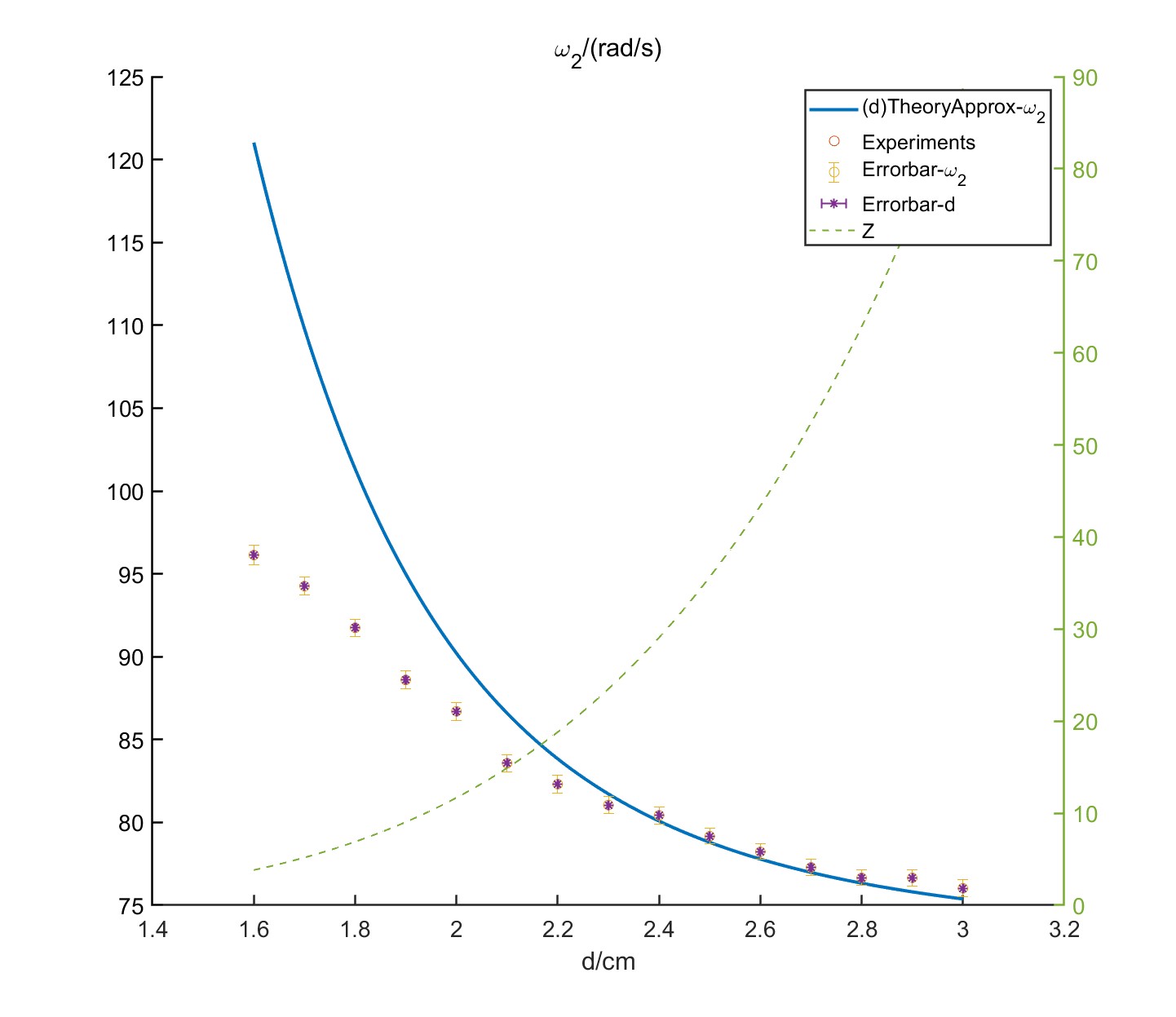}}
\end{minipage}

\caption{\textbf{Peak frequency changing with the independent variable.} 
{\small These data values were obtained by performing an FFT operation on the time-domain data. (a) and (b) correspond to the experimental group where only the plate length was varied.(c) and (d) represent the experimental group where only the distance between the plates was varied. From these figures, we can see that the peak value decreases as the plate length increases, and under the same conditions, the peak frequency of anti-phase motion is higher than that of in-phase motion. With plate length fixed, as the distance between the plates increases, the peak frequencies of anti-phase and in-phase motions tend to be identical.}}
\label{Fig.15}
\end{figure}

Performing FFT (Fast Fourier Transform) on time-domain data, the peak frequency is extracted from the frequency domain as the data value. Repeating the experiment multiple times, we obtained the results of the peak frequency 
changing with the independent variable, as shown in \textbf{Fig.\ref{Fig.15}}.

\textbf{Figures \ref{Fig.15(a)}} and \textbf{\ref{Fig.15(b)}} correspond to the experimental group where only the plate length is varied. From these, we can observe some similarities and differences between 
$\omega_1$ and $\omega_2$:

1.Similarities: As the plate length $L$ increases, the peak values of both frequencies show a downward trend. This is because the increase in plate length leads to a decrease in the effective stiffness coefficient and a slight increase in mass.

2.Differences: Under the same conditions, $\omega_2$ is always greater than $\omega_1$ . This is because the repulsive force contribution from the magnetic interaction exists in the governing Eq. (\ref{Func.2.10(b)}) for the antisymmetric mode. This repulsive force decreases with increasing distance and acts similarly to a restoring force.

\vspace{7pt}
\textbf{Figures \ref{Fig.15(c)}} and \textbf{\ref{Fig.15(d)}} correspond to the experimental group where only the distance between the plates is varied. Under these experimental conditions, there is a significant difference between $\omega_1$ and $\omega_2$ .

In \textbf{Fig.\ref{Fig.15(c)}}, the theoretically predicted angular frequency $\omega_1$ for the symmetric mode is approximately equal to the angular frequency of a single plate vibrating independently, and it is not related to the interaction force between the magnets. Therefore, we see in \textbf{Fig.\ref{Fig.15(c)}} that when only d is changed, $\omega_1$ remains almost unchanged and is close to our theoretical prediction.

The difference between $\omega_1$ and $\omega_1$is caused by the interaction force between the magnets. Thus, we can see that as the distance d between the plates increases, the strength of the interaction force between the magnets weakens, and the difference between the two modes naturally diminishes.

\vspace{7pt}
Throughout the experiment, we found that the experimental values were slightly lower than the theoretical predictions, and the simulation-calculated $K_e$ was slightly higher than the theoretical value. We speculate that this discrepancy may arise from air damping and some deviations in the effective mass $M_{eff}$ . Besides translational kinetic energy, the magnets also have rotational kinetic energy, and higher-order vibration modes of the plates contribute some kinetic energy as well. Since $M_{eff}$ does not account for this additional kinetic energy, it is slightly underestimated. However, the current simplified model is already sufficiently refined.

\vspace{7pt}
Overall, from the experimental results, we see that the theoretical value of the normal frequency $\omega_1$ corresponding to the linear Eq. (\ref{Func.2.10(a)}) matches well with the experimental value, while the normal frequency 
$\omega_2$ corresponding to the nonlinear Eq. (\ref{Func.2.10(b)})shows a clear deviation from the theoretical approximation. This deviation is more pronounced when d is smaller (stronger magnetic force) or L is larger (weaker elastic force).

In \textbf{Fig.\ref{Fig.15(b)}} and \textbf{Fig.\ref{Fig.15(d)}}, we also show the curve of the previously defined $Z$ value with respect to the experimental independent variables, which characterize the relative strength of the elastic force to the magnetic force. By comparison, we can see that when $Z>20$ which means elastic force is much larger than magnetic force, the deviation does not exceed $5\%$, indicating that under this condition, the elastic force is dominant, and our theory is sufficiently applicable.

\section{Didactic considerations}

The analytical approach and the use of simulation software in our analysis can also be applied to other aspects of daily life and teaching.

The magnetic mechanical oscillator is a simple dual-degree-of-freedom vibrational system that can be constructed at home or in the classroom. 
It serves as an excellent teaching case for demonstrating multi-degree-of-freedom vibrations and simple nonlinear effects. Its key components, 
such as the vibration of a single plate, the magnetic field distribution generated by the magnet, and the forces between two magnets, can be used as effective teaching demonstrations. The article provides details about the experimental equipment and data used for quantitative experiments and The COMSOL simulation files that we have developed are available for free on GitHub:\href{https://github.com/quantumopticss/MMO}{https://github.com/quantumopticss/MMO}

We believe that the theoretical and experimental studies presented in this work will be valuable for understanding relevant phenomena.

\section{Conclusion}
This article investigates a dual-degree-of-freedom vibrational system composed of magnets and a vibrating plate from the 
perspective of vector mechanics and differential equations.

The study demonstrates the coupled vibrational behavior of two plate-like spring blades under magnetic force interaction. 
Initially, we attempt to qualitatively understand the magnetic mechanical oscillator system: each plate can vibrate independently, 
and the vibration of one plate induces changes in the magnetic force generated by the magnet. Thus, under magnetic force coupling, 
the vibrations of the two plates become correlated, forming a dual-degree-of-freedom vibrational system.

To gain a deeper understanding of the two normal modes of the magnetic mechanical oscillator, we employ a modular approach to model 
and analyze the properties of each part of the system. We start by establishing a model for analyzing the vibration of a single plate 
based on the Yang's equation. Subsequently, we analyze the experimental environment, model the magnetic field produced by the magnet, 
and examine the forces acting on the magnet. From the dynamic equations of the system, we deduce various properties of the system's motion: 
the magnetic mechanical oscillator is a dual-degree-of-freedom normal system, divided into in-phase and out-of-phase vibration modes. 
The theoretical predictions for the frequencies of these two modes are provided. We introduce an indicator, $Z$, to measure the relative 
magnitudes of elastic and magnetic forces. Our results show that when $Z>20$, elastic forces dominate over magnetic forces, aligning well 
with experimental observations.

For the various approximations used in our modeling and theoretical analysis, we employ the Comsol simulation software to supplement the 
validation of theoretical reasonability and feasibility ranges. Finally, we experimentally validate our theoretical predictions.

It is worth noting that in this research, we effectively modularized the study of complex problems, achieving simplicity while maintaining good experimental agreement. Therefore, our goal is not to consider and precisely discuss every minor factor (which is, in fact, impossible), but rather to achieve excellent results overall through this modular approach. The deviations in the nonlinear parts, where detailed discussion is not overly specific, remain small.

Simplifying complex problems while achieving excellent fits is the highlight of our study.

\section*{Appendix}

\setcounter{table}{0}
\setcounter{figure}{0}
\gdef\thesection{Appendix \Alph{section}}
\gdef\thesubsection{\Alph{section}.\arabic{subsection}}
\gdef\theequation{\Alph{section}.\arabic{equation}}

\section{Detailed analysis process of the plate}

\begin{figure}[h]
\centering
\begin{minipage}[t]{0.48\textwidth}
\centering
\includegraphics[width=1\textwidth]{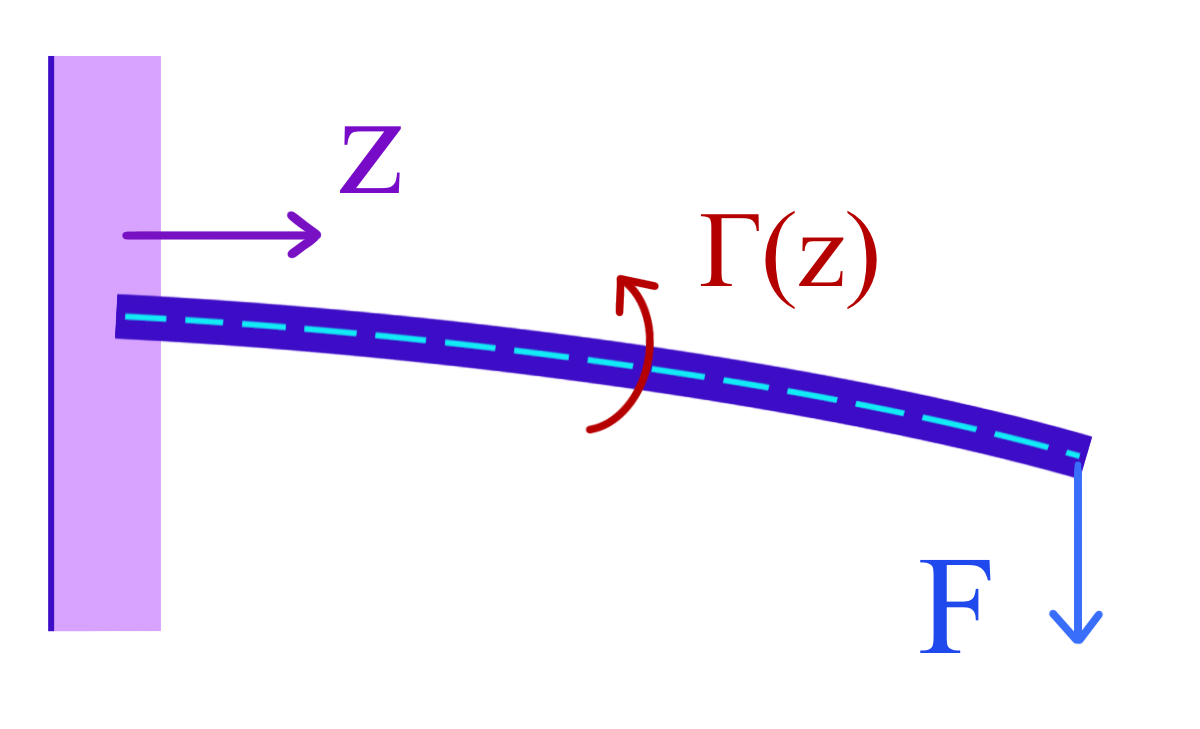}
\caption{\textbf{Torque balance diagram of a curved plate with small deflection.} T{\small he moment $\Gamma$ transmitted due to bending at the cross-section, the stress $F$ applied, and the reference position coordinate $Z$.}}
\label{Fig.A1}
\end{minipage}
\hspace{9pt}
\begin{minipage}[t]{0.48\textwidth}
\centering
\includegraphics[width=1\textwidth]{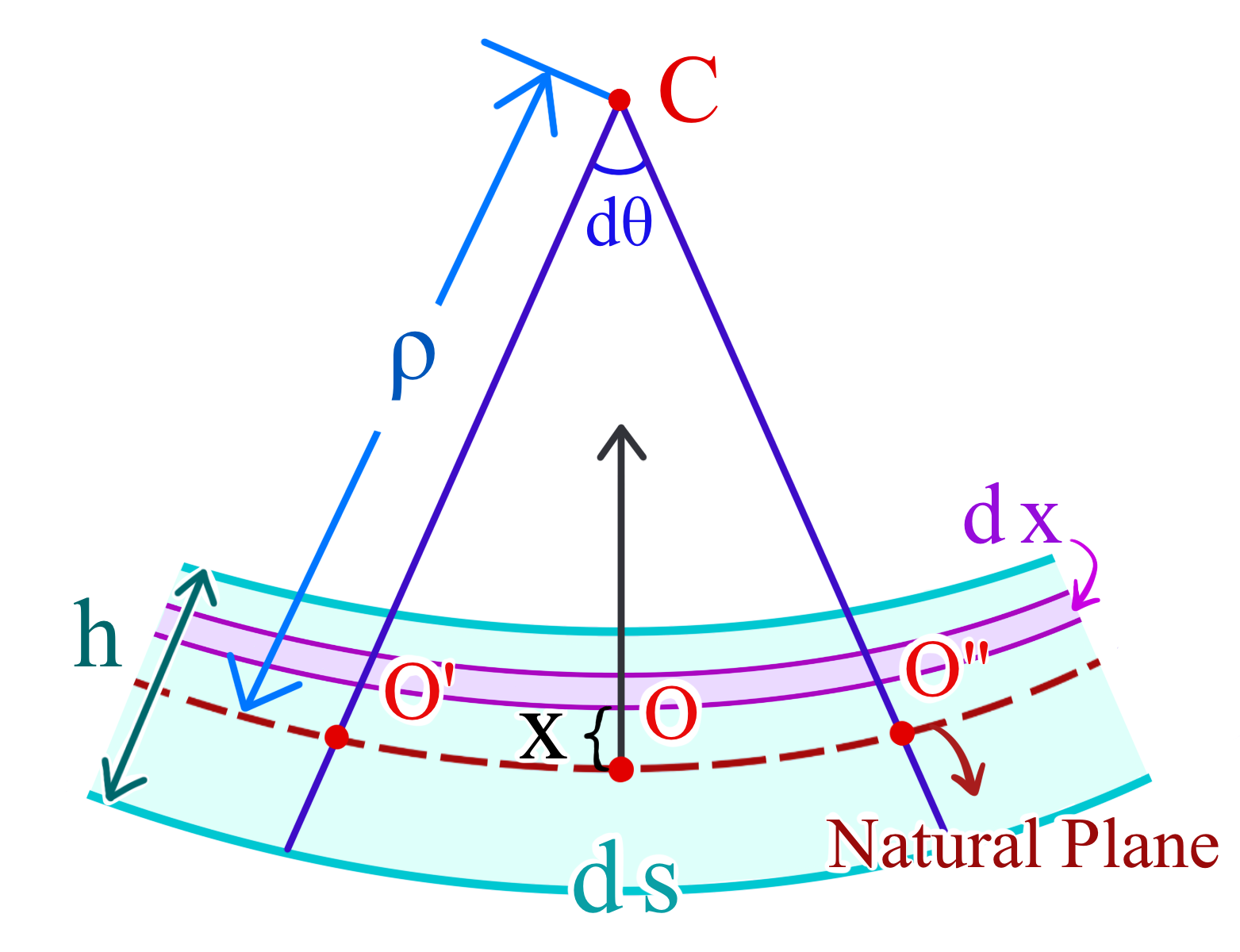}
\caption{\textbf{Details in position z.} {\small From this, we can geometrically analyze and obtain the required physical quantities at that location. At this point, we are considering a plate undergoing small deflection bending.}}
\label{Fig.A2}
\end{minipage}
\end{figure}

We abstract the plate-shaped leaf spring as a classical spring and treat the magnet and the spring as a whole with an equivalent mass. Based on laboratory observations and its favorable geometric properties, we consider the small deflection bending of the plate as shown in \textbf{Fig.\ref{Fig.A1}}.

\textbf{Figure \ref{Fig.A2}} shows the details of the cross-section of the plate with small deflection bending. From the geometry, we can derive the following:

\begin{equation}
  \begin{aligned}
    &\rho = \displaystyle\frac{d s}{d \theta} \hspace{1cm} \displaystyle\frac{\Delta L(x)}{L} \approx \displaystyle\frac{x}{\rho}  \hspace{1cm} \text{where x $\in [-\frac{h}{2},\frac{h}{2}]$}
  \end{aligned}
\end{equation}

The stress-strain relationship of a linearly elastic solid follows Hooke's Law, and if the Poisson's ratio effect is ignored, we have:

\begin{equation}
  \begin{aligned}
    &F = YA\displaystyle\frac{\Delta L}{L} \approx YA\displaystyle\frac{x}{\rho}
  \end{aligned}
\end{equation}

The bending moment transmitted by the force on a cross-section of the plate can be calculated by integrating:

\begin{equation}
  \begin{aligned}
    &\Gamma = \int_{-\frac{h}{2}}^{\frac{h}{2}} Ya \frac{x}{\rho}x \,dx = \displaystyle\frac{Yah^3}{12 \rho}
  \end{aligned}
\end{equation}

We apply a load $F$ at the end of the plate, and the bent plate should satisfy the moment equilibrium equation:

\begin{equation}
  \begin{aligned}
    &\Gamma (z) = \displaystyle\frac{Yah^3}{12 \rho(z)} \approx F(L-z)
  \end{aligned}
  \label{Func.A.4}
\end{equation}

In the small deflection approximation, the radius of curvature satisfies a very good approximation:

\begin{equation}
  \begin{aligned}
    &\rho = \displaystyle\frac{(1+{y^{'}}^{2})^{\frac{3}{2}}}{y^{''}} \approx \displaystyle\frac{1}{y^{''}}
  \end{aligned}
\end{equation}

Substituting it into Eq. (\ref{Func.A.4}) and simplifying, we will get:

\begin{equation}
  \begin{aligned}
    &y^{''} \approx \displaystyle\frac{12F(L-z)}{Yah^3}
  \end{aligned}
  \label{Func.A.6}
\end{equation}

Integrate Eq. (\ref{Func.A.6}) and substitute boundary conditions: $y|_{z=0}=0 \hspace{0.1cm}\displaystyle\frac{dy}{dz}|_{z=0}=0$, 
we will get the equation of $y(z)$, which shows that even in the case of small deformation, the bending of the plate is still non-linear, and we should be careful when handling the kinetic energy of the bending plate vibration process.

\begin{equation}\label{yx}
  \begin{aligned}
    &y(z) = \displaystyle\frac{12 F}{Yah^3} (\frac{1}{2} Lz^{2} - \frac{1}{6}z^3)
  \end{aligned}
\end{equation}

Considering the displacement at the end of the spring under the load, \(y_{\text{end}} = y|_{z=L} = \frac{4 FL^3}{Yah^3}\), as the deformation produced by the load \(F\) applied to an equivalent spring, we obtain that the force applied to the plate is proportional to the linear restoring force generated by the deformation at the end: \(F_e = -K_e x\) (the negative sign indicates that the restoring force is opposite to the direction of deformation). Therefore, the expression for the equivalent stiffness coefficient\(K_e\) is:

\begin{equation}
  \begin{aligned}
    &K_e = \displaystyle\frac{F}{y_{end}} = \displaystyle\frac{Yah^3}{4L^3} 
  \end{aligned}
\end{equation}

If considering the Poisson's ratio effect on the deformation of the plate, the nature of the linear restoring force remains unchanged. However, it necessitates a modification in the expression of the equivalent stiffness coefficient $K_e$ \footnote[7]{For more details, please refer to reference \cite{item12}}:

\begin{equation}
  \begin{aligned}
    &K_e = \displaystyle\frac{Yah^3}{4(1-\nu^2)L^3} 
  \end{aligned}
\end{equation}

Now, we start from an energy perspective to calculate the equivalent mass of the system. Under the small deflection approximation, the deformation pattern of the plate is nearly identical to that in the static state. Therefore, it can be assumed that the vibration equation of the plate under small amplitude conditions satisfies quasi-static conditions, i.e., the vibration of the plate is dominated by the fundamental mode. Deforming 
Eq. (\ref{yx}) and assuming, the following conditions rise:

\begin{equation}
  \begin{aligned}
    y(z,t) \approx \displaystyle\frac{3y_{end}(t)}{L^3}\left(\displaystyle\frac{Lz^2}{2} - \displaystyle\frac{z^3}{6}\right)
  \end{aligned}
\end{equation}

On this basis, the energy of the plate when the end velocity is \(\dot{y_{\text{end}}}\) is calculated as:

\begin{equation}
  \begin{aligned}
    &E_k = \int_{0}^{L} \frac{1}{2} \rho_e a h \left(\displaystyle\frac{\partial y(z,t) }{\partial t} \right)^{2} \,dz = \frac{1}{2} M_{e} (\dot{y_{end}})^{2} 
    \\
    &M_{e} = \alpha \rho_e ahL \hspace{1cm} \alpha = 9(\frac{1}{20}+\frac{1}{252}-\frac{1}{36})
  \end{aligned}
\end{equation}

Taking into account the magnet fixed at the end of the plate, the expression for 
$M_{eff}$ is obtained as:

\begin{equation}
  \begin{aligned}
    &M_{eff} = \alpha \rho_e ahL + m_0
  \end{aligned}
\end{equation}

In summary, we analyzed that the spring force $F_e$ of the plate is a linear restoring force: $F_e = -K_e x$ and provided the expressions for the equivalent stiffness $K_e$ and the equivalent mass $M_{eff}$.

\section{Detailed analysis process of the magnets}

In this section, we will analyze the form of the interaction forces between magnets. We can approximate the magnetic mechanical oscillator system to 
be in an environment with $\mu = \mu_0 \hspace{0.2cm} \epsilon = \epsilon_0$, and devoid of free charge $\rho_f$ and free current $\mathbf{j_f}$ ($\sigma \approx 0$) in the surroundings.
\\
We list Maxwell's equations as follows:

\begin{equation}
  \begin{aligned}
    &\nabla \bullet \mathbf{D} = 0 \hspace{1cm} \nabla \times \mathbf{E} = -\displaystyle\frac{\partial \mathbf{B}}{\partial t}
    \\
    &\nabla \bullet \mathbf{B} = 0 \hspace{1.03cm} \nabla \times \mathbf{H} = \displaystyle\frac{\partial \mathbf{D}}{\partial t}
  \end{aligned}
\end{equation}

We can safely neglect the radiation effects of the system, approximating it to satisfy 
\(\nabla\times\mathbf{H} = 0\). We can use the magnetic scalar potential \(\varphi_m\) to describe the magnetic field around the magnetic mechanical oscillator system: 
\(\mathbf{H} = -\nabla\varphi_m\).

Next, we introduce the magnetic charge theory for the magnetic field. By combining 
\(\nabla\cdot\mathbf{B} = 0\) and \(\mathbf{B} = \mu_0(\mathbf{H}+\mathbf{M})\), we obtain:
\begin{equation}
  \nabla\cdot\mathbf{H} = -\nabla\cdot\mathbf{M}
\end{equation}
Mathematically, the spatial magnetic charge density can be considered as:

\begin{equation}
  \rho_m = -\nabla\cdot(\mu_0\mathbf{M})
\end{equation}

The surface density of the upper magnetic charge on the boundary surface is given by:

\begin{equation}
  \begin{aligned}
    &\sigma_m = \mu_0 \mathbf{\hat{n_{12}}} \cdot \left[(\mathbf{M_1} - \mathbf{M_2})\right] 
  \end{aligned}
\end{equation}

Therefore, for a uniformly magnetized cylindrical magnet with magnetization intensity $\mathbf{M}$ 
parallel to its axis. It is equivalent to a magnetic dipole layer with a surface magnetic charge density of $\sigma_m$. 
The diameter of the dipole layer is the same as the magnet diameter $D$, and the thickness of the 
magnet corresponds to the distance between the magnetic dipole layers, denoted as $W$.

\begin{figure}[h]
\centering
\begin{minipage}[t]{0.48\textwidth}
\centering
\includegraphics[width=1\textwidth]{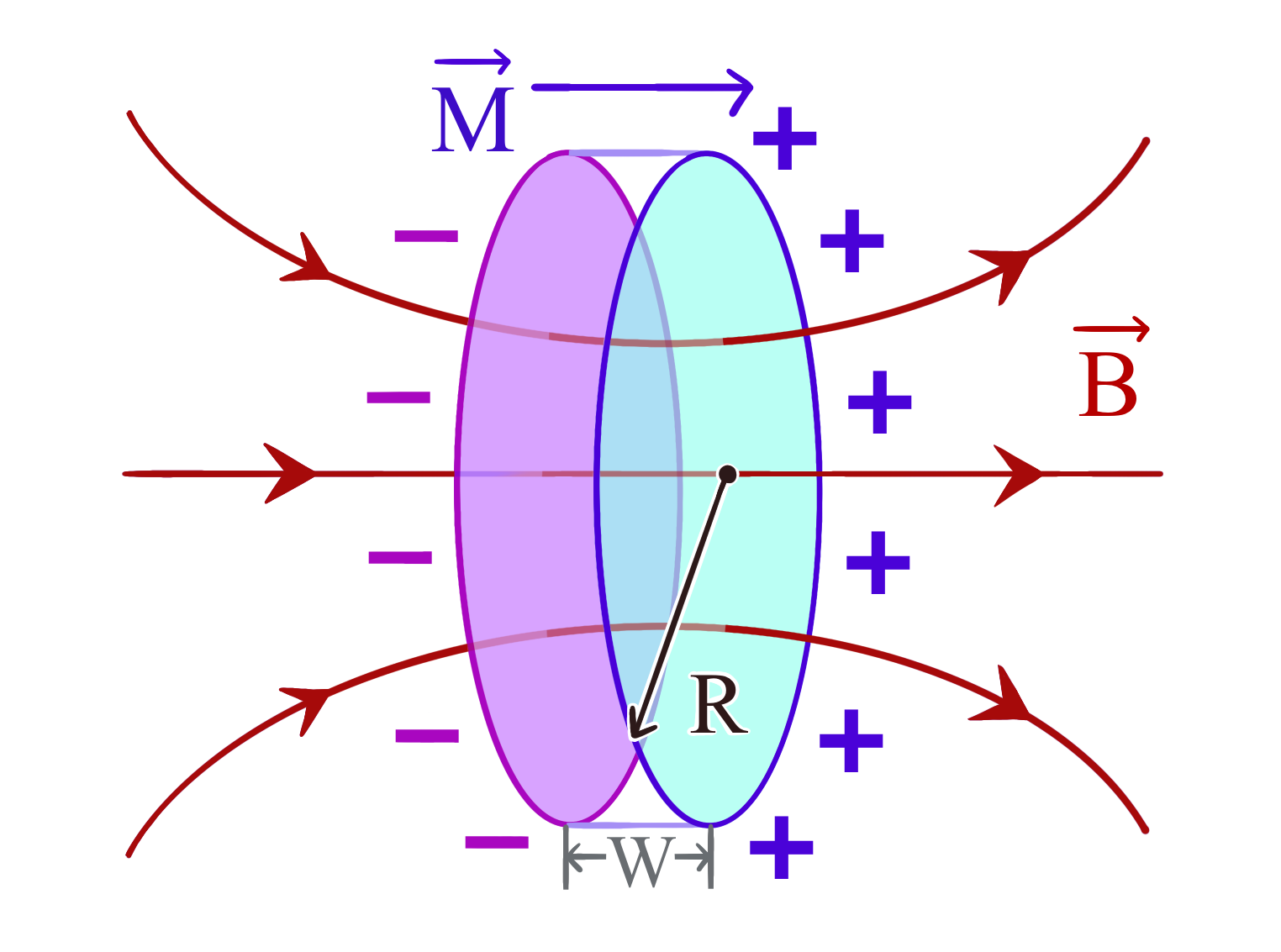}
\caption{\textbf{Diagram of a magnet.} {\small This is  a magnetic charge distributed on a geometric circle with a diameter of D,  with its axis along the z-axis.}}
\label{Fig.B1}
\end{minipage}
\hspace{9pt}
\begin{minipage}[t]{0.48\textwidth}
\centering
\includegraphics[width=1\textwidth]{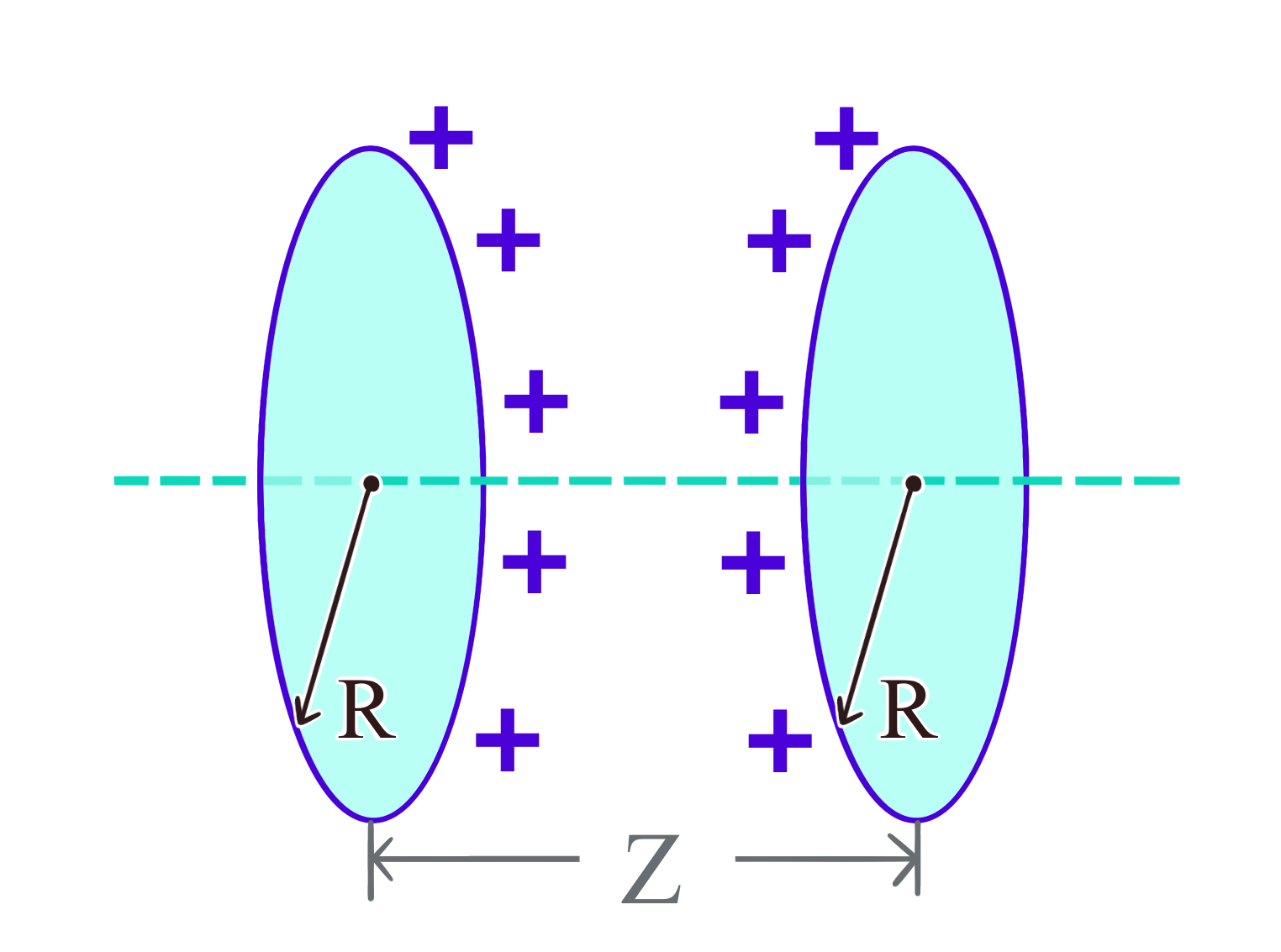}
\caption{\textbf{Two layers of magnetic charges.} {\small We consider two identical parallel and coaxially placed surface-distributed magnetic charges.}}
\label{Fig.B2}
\end{minipage}
\end{figure}

For a magnetic charge distributed on a geometric circle with a diameter of $D$, as illustrated in \textbf{Fig.\ref{Fig.B1}}, 
with its axis along the $z$-axis, in cylindrical coordinates, the magnetic potential at the point $(\rho, \varphi, z)$ 
is given by:

\begin{equation}\label{phim}
  \begin{aligned}
    &\varphi_{m0} = \int_{0}^{\frac{D}{2}} r \,dr \int_{0}^{2\pi} \,d\phi \displaystyle\frac{\sigma_m}{4\pi\mu_0 \sqrt{z^2 + \rho^2 + r^2 -2r\rho\cos{\phi}}}
  \end{aligned}
\end{equation}

Due to the small deformation of the plates, we can approximate that the two plates always remain 
parallel to each other. Therefore, considering two identical parallel and coaxially placed surface 
distributed magnetic charges, as illustrated in \textbf{Fig.\ref{Fig.B2}}, the calculation of the magnetic interaction energy between them is given by:

\begin{equation}
  \begin{aligned}
    E(z) = 2\pi\int_{0}^{\frac{D}{2}} r_2 \,dr_2 \int_{0}^{\frac{D}{2}} r_1 \,dr_1 \int_{0}^{2\pi} \,d\phi \displaystyle\frac{\sigma_m^2}{4\pi\mu_0 \sqrt{z^2 + r_1^2 + r_2^2 -2r_1r_2\cos{\phi}}}
  \end{aligned}
\end{equation}

For two magnetic pieces at a distance of $z$, considering all surfaces is necessary. Therefore, the total mutual interaction energy is given by:

\begin{equation}
  \begin{aligned}
    E(z) &= E_{++}(z-W) + E_{--}(z+W) + E_{+-}(z) + E_{-+}(z)
  \end{aligned}
\end{equation}

Taking the gradient will yield the mutual interaction force $F_m(z)$ between them:

\begin{equation}
  \begin{aligned}
    &F_m(z) = - \displaystyle\frac{d E(z)}{dz} \approx \displaystyle\frac{3\mu_0 m^2}{2\pi z^4} + \displaystyle\frac{5\mu_0 m^2(W^2 - \frac{3D^2}{4})}{4\pi z^6} 
  \end{aligned}
\end{equation}

where $m$ is the magnetic moment:

\begin{equation}
  \begin{aligned}
    &\mathbf{m} = \mathbf{M}V \hspace{0.5cm} V = \displaystyle\frac{\pi D^2 W}{4}
  \end{aligned}
\end{equation}

In summary, we obtain the equation governing the approximate force $\mathbf{F_m}$ between two magnets in the magnetic-mechanical oscillator system
\footnote[7]{If the distance between the two plates increase, the approximation presented in reference \cite{item5} can be employed, considering only the term $K_{m1}$ in the magnetic force.}.

\begin{equation}
  \begin{aligned}
    &F_m(z) = \displaystyle\frac{K_{m1}}{z^4} + \displaystyle\frac{K_{m2}}{z^6}
    \\
    &K_{m1} = \displaystyle\frac{3\mu_0 {m^2}}{2\pi} \hspace{0.5cm} K_{m2} = \displaystyle\frac{5\mu_0 m^2 (W^2 - \frac{3D^2}{4})}{4\pi}
  \end{aligned}
\end{equation}

\section{Detailed process of solving the differential equations}

Here, we will separately solve the two differential equations corresponding to the symmetric and antisymmetric modes of the blade spring movement in the main text Eqs. (\ref{Func.2.10(a)}) and (\ref{Func.2.10(b)}).

\begin{figure}[h]
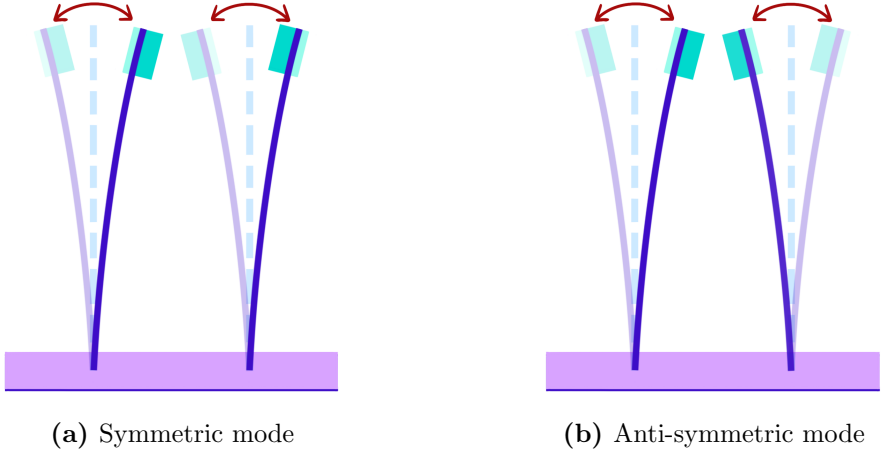

\begin{minipage}[b]{0.45\linewidth}
    \centering
    \subfloat[Symmetric mode]{\label{Fig.C1(a)}\includegraphics[width=0.95\linewidth]{figures/sym_mode.png}}
\end{minipage} 
\medskip
\begin{minipage}[b]{0.45\linewidth}
    \centering
    \subfloat[Anti-symmetric mode]{\label{Fig.C1(b)}\includegraphics[width=0.95\linewidth]{figures/anti_mode.png}}
\end{minipage}

\caption{\textbf{The two modes of the blade spring's motion.} {\small (a) can be solved by Eq. (\ref{Func.2.10(a)}) and (b) can be solved by Eq. (\ref{Func.2.10(b)}).}}
\label{Fig.C1}
\end{figure}

\begin{subequations}
  \begin{equation}
    \ddot{y_1} = \displaystyle\frac{-\beta \dot{y_1}}{M_{eff}} - \displaystyle\frac{K_e y_1}{M_{eff}}
  \label{Func.C.1a}
  \end{equation}
  \begin{equation}
    \ddot{y_2} = \displaystyle\frac{-\beta \dot{y_2}}{M_{eff}} - \displaystyle\frac{K_e y_2}{M_{eff}} - \displaystyle\frac{2K_{m1}}{M_{eff}(d - y_2)^4} - \displaystyle\frac{2K_{m2}}{M_{eff}(d - y_2)^6}
  \label{Func.C.1b}
  \end{equation}
\end{subequations}

Equation (\ref{Func.C.1a}) is a simple second-order ordinary differential equation with constant coefficients. By using the initial 
conditions \(y_1|_{t=0}\) and \(\dot{y_1}|_{t=0}\), we can solve for:

\begin{equation}
\begin{aligned}
y_1(t) = e^{-\displaystyle\frac{\beta t}{2M_{eff}}}\left[ y_1|_{t=0}\cos{\omega_1 t} + \displaystyle\frac{\dot{y_1}|_{t=0}}{\omega_1}\sin{\omega_1 t}\right]
\end{aligned}
\end{equation}

where:

\begin{equation}
  \omega_1 = \sqrt{\displaystyle\frac{K_e}{M_{eff}} - \displaystyle\frac{\beta^2}{4M_{eff}^2}} \approx \sqrt{\displaystyle\frac{K_e}{M_{eff}}}
\end{equation}

Equation (\ref{Func.C.1b}) is a nonlinear differential equation, and we can use numerical methods to solve it. However, it's worth noting that we can still effectively handle it through substitution and perturbation methods. When \(y_2 \ll d\) and the elastic force is relatively stronger than the magnetic force, we can approximately simplify Eq. (\ref{Func.C.1b}):

\begin{equation}\label{y2}
  \begin{aligned}
  \ddot{y_2} \approx \displaystyle\frac{-\beta \dot{y_2}}{M_{eff}} - \displaystyle\frac{K_e y_2}{M_{eff}} &- \displaystyle\frac{2K_{m1}}{M_{eff}d^4} \left[1 + \displaystyle\frac{4 y_2}{d} + 10\left(\displaystyle\frac{y_2}{d}\right)^2 + 20\left( \displaystyle\frac{y_2}{d} \right)^3 + o\left( \displaystyle\frac{y_2}{d} \right)^4 \right]
  \\
  &- \displaystyle\frac{2K_{m2}}{M_{eff}d^6} \left[1 + \displaystyle\frac{6 y_2}{d} + 21\left(\displaystyle\frac{y_2}{d}\right)^2 + 56\left( \displaystyle\frac{y_2}{d} \right)^3 + o\left( \displaystyle\frac{y_2}{d} \right)^4 \right]
  \end{aligned}
\end{equation}

Denote:

\begin{equation}
  \begin{aligned}
    \omega_{20} &= \sqrt{\displaystyle\frac{K_e}{M_eff} + \displaystyle\frac{8K_{m1}}{M_{eff}d^5} +\displaystyle\frac{12K_{m2}}{M_{eff}d^7}} \hspace{0.7cm} \alpha = \displaystyle\frac{20K_{m1}}{M_{eff}d^6} + \displaystyle\frac{42K_{m2}}{M_{eff}d^8}
    \\
    \beta &= \displaystyle\frac{40K_{m1}}{M_{eff}d^7} + \displaystyle\frac{112K_{m2}}{M_{eff}d^9} \hspace{2.7cm} \theta = \displaystyle\frac{2K_m}{M_{eff}d^4} + \displaystyle\frac{2K_{m2}}{M_{eff}d^6}
  \end{aligned}
\end{equation}

We have:

\begin{equation}\label{y2}
  \begin{aligned}
  &\ddot{y_2} + \displaystyle\frac{\beta \dot{y_2}}{M_{eff}} +  \omega_{20}^2 y_2 \approx -\theta  - \alpha y_2^{2} - \beta y_2^{3}
  \end{aligned}
\end{equation}
Let $z_2 = y_2 + \displaystyle\frac{\theta}{\omega_{20}^2}$, we will get:
\begin{equation}\label{y2}
  \begin{aligned}
  &\ddot{z_2} + \displaystyle\frac{\beta \dot{z_2}}{M_{eff}} + \left[\omega_{20}^2 -\displaystyle\frac{2\alpha\theta}{\omega_{20}^2} +\displaystyle\frac{3\beta \theta^2}{\omega_{20}^{4}} \right] z_2 \approx   - (\alpha -\displaystyle\frac{3\beta \theta}{\omega_{20}^2}) z_2^{2} - \beta z_2^{3}
  \end{aligned}
\end{equation}

The normal mode \(z_2\) is approximated as a standard nonlinear vibration with an amplitude \(A\) and angular frequency \(\omega_2\). Landau's theoretical mechanics textbook, citation \cite{item12}, provides some key conclusions:

\begin{align}
    &\Omega = \sqrt{\omega_{20}^2 -\displaystyle\frac{2\alpha\theta}{\omega_{20}^2} +\displaystyle\frac{3\beta \theta^2}{\omega_{20}^{4}}}
    \\
    &\omega_2 \approx \Omega + \left[ \displaystyle\frac{3\beta}{8\Omega} - \displaystyle\frac{5\left(\alpha - \frac{3\beta\theta}{\omega_{20}^2} \right)^2}{12\Omega^3} \right]A^2
  \end{align}

For the qualitative comparison mentioned earlier, where elastic force is relatively stronger than magnetic force, we define a dimensionless number $Z$ to characterize the relative strength of the two:

\begin{equation}
  Z = \displaystyle\frac{K_e d^5}{K_{m1}}
\end{equation}

We will demonstrate in the experimental section that \(Z\) can be used to determine under what conditions our theory is reasonable.

\printbibliography

\end{document}